\documentclass[9pt,twocolumn,twoside]{gsajnl}

\usepackage{amsmath,amssymb,graphicx,bm}
\usepackage{hyperref}
\usepackage{xcolor}
\usepackage[percent]{overpic}
\usepackage{booktabs}
\usepackage{comment}
\usepackage{epstopdf}
\usepackage{changepage}

\articletype{inv} 

\DeclareSymbolFont{matha}{OML}{txmi}{m}{it}
\DeclareMathSymbol{\varv}{\mathord}{matha}{118}

\newcommand{\fix}{\mathrm{fix}}
\newcommand{\avg}[1]{\left\langle #1\right\rangle}
\newcommand{\Dop}{\mathcal D}
\newcommand{\Sop}{\mathcal S}
\newcommand{\Mop}{\mathcal M}
\newcommand{\Lop}{\mathcal L}

\runningtitle{Multi-allele fixation probabilities} 
\runningauthor{Yuzbashyan \textit{et al.}}

\title{Fluctuation theorems and exact multi-allele fixation probabilities}

\author[1]{Emil A. Yuzbashyan}
\author[2]{Alexander Feigel}
\author[3,$\ast$]{Alexandre V. Morozov}

\affil[1]{Department of Physics and Astronomy, Center for Materials Theory, Rutgers, The State University of New Jersey, 136 Frelinghuysen Road, Piscataway, New Jersey 08854, USA}
\affil[2]{The Racah Institute of Physics, The Hebrew University of Jerusalem, Edmond J. Safra Campus, Jerusalem 9190401, Israel}
\affil[3]{Department of Physics and Astronomy, Rutgers, The State University of New Jersey, 136 Frelinghuysen Road, Piscataway, New Jersey 08854, USA}

\correspondingauthoraffiliation[$\ast$]{Corresponding author: morozov@physics.rutgers.edu.}

\begin{abstract}
A mutant gene that appears in a finite population will, in the long run, either disappear from it or become fixed, so that all members of the population will carry a copy of the mutant gene.
Determining the fixation probability of a mutant gene is thus a prerequisite for modeling evolutionary dynamics in finite populations.
Within the diffusion theory framework, the mutant's competition with the wild-type genes is described by the closed-form solution of a one-dimensional backward Kolmogorov equation
for the mutant fixation probability, derived by Motoo Kimura in 1962.
However, in asexual populations the rate of adaptation and the relative importance of alternative evolutionary pathways are shaped by clonal interference -- the competition between several lineages arising
from distinct beneficial or deleterious mutations.
Quantitative descriptions of clonal interference require computing fixation probabilities in populations where multiple alleles are present simultaneously -- a formidable mathematical problem
which amounts to solving a multi-dimensional backward Kolmogorov equation.
Here, we employ a path integral approach to fluctuation theorems (a set of exact equalities in non-equilibrium statistical mechanics) in order to re-derive Kimura's fixation
probability formula. When extended to multiple-allele systems, fluctuation theorems provide a constraint which must be satisfied by any valid set of fixation probabilities. Moreover, we
present a recursive analytical solution of the multiple-allele fixation probability problem. We use our formalism to study the fate of a novel beneficial mutation in a clonal interference scenario,
discovering substantial deviations between our exact treatment and the mean-field approximation in which the target mutation competes with a single effective genotype assigned an average of the selection
coefficients of multiple interfering mutations that are actually present in the population.
Our formalism can be used to replace both explicit simulations and approximate analytical results in population genetics models of asexual evolution.
\end{abstract}

\keywords{Fixation probabilities; Fluctuation theorems; Multiple alleles; Clonal interference; Asexual evolution; Bacteria}

\begin{document}




\maketitle



\section{Introduction}

Evolution of biological populations is subject to three major forces: mutation, natural selection, and migration~\citep{Wright1931,Fisher1958}.
Besides these deterministic forces, finite populations undergo random genetic drift -- stochasticity that results, in mathematical models of evolution,
from probabilistic sampling of individuals from one generation to the next~\citep{Kimura1971,Kimura1983,Ewens2004,Gillespie2004}. Genetic drift
reflects randomness of survival or reproduction in biological populations; it can dominate the adaptive effects of natural selection when the population size is small.

Evolution proceeds by creating new deleterious, neutral, or beneficial mutant alleles. At first, these new alleles are present in a single copy within a population -- as a result, the fate of mutant alleles
is subject to stochastic effects. In finite-size populations with very low mutation rates (so that different mutants do not interfere with one another),
each mutant allele will be either lost or fixed in the population before the next mutation arises~\citep{Kimura1983,Crow2009}. Fixation means that the mutation will be carried by all members of the population;
however, even beneficial mutations that would have conferred a selective advantage can disappear from the population due to random chance.

The predominant quantitative framework for computing fixation probabilities is based on the diffusion approximation, where both allele frequencies and time are treated as continuous variables~\citep{Kimura1964,Kimura1983,Ewens2004}. Within this framework, the probability of fixation of a single mutant arising on the homogenous wild-type background in a well-mixed
population is given by the celebrated Kimura's formula~\citep{Kimura1962}. However, there can be more than two segregating alleles at a single protein or nucleic acid site.
For example, viral proteins evolve rapidly under the selective pressure of antiretroviral drugs, leading to multiple alternative drug-resistance mutations at the same site~\citep{Rhee2007}.
The T69 locus in HIV-1 reverse transcriptase can mutate to aspartate (T69D), asparagine (T69N), or serine (T69S) in different patients, conferring varying levels of resistance to nucleoside reverse transcriptase inhibitors.
Similarly, the T215 locus is a mutational hotspot that can undergo tyrosine (T215Y), phenylalanine (T215F), or other amino acid substitutions depending on the antiviral treatment regimen.
Thus, exact computation of fixation probabilities for multiple simultaneously competing alleles would be of considerable theoretical and practical interest. 

Moreover, multi-allele fixation probabilities lie at the heart of the studies of genome-wide, multi-locus evolution in the absence of recombination.
Indeed, even if the mutation rate per locus, $\mu$, is small, the effective mutation rate per genome,
$\mu L$, may be large enough to allow for the simultaneous appearance of multiple beneficial mutations in the population (here, $L$ is the number of non-conserved amino acids or nucleotides).
Clonal interference between distinct lineages that are simultaneously present in asexual populations 
has received considerable attention in theoretical population genetics~\citep{Tsimring1996,Gerrish1998,Rouzine2003,Wilke2004,Park2007,Desai2007a,Desai2007b,Fogle2008,Park2010,Schiffels2011,Hallatschek2011,Good2012}.
These works typically model mutation-selection balance in evolving asexual populations,
focusing on the speed and degree of adaptation and on the fate of individual mutations in the presence of genetic linkage. Some of these models rely on approximations to multi-allele fixation probabilities -- although the exact
solution of the two-allele case has been available since 1962~\citep{Kimura1962} and is routinely explained in textbooks~\citep{Kimura1983,Gillespie2004,Crow2009}, its extension to three or more alleles has proven difficult,
requiring various approximations and empirical simplifications~\citep{Lessard2009,Overall2024,Braga2026}.

In this work, we establish a connection between fluctuation theorems (exact relations satisfied by out-of-equilibrium systems in statistical mechanics) and fixation probabilities in population genetics.
We use this approach to re-derive Kimura's formula for the two-allele system and obtain a constraint satisfied by fixation probabilities in the multiple-allele case.
Guided by the fluctuation theorem constraint, we derive the exact solution of the multi-allele backward Kolmogorov equation, valid in weak, intermediate, and strong selection regimes.
We check our solution against numerical results, establishing its consistency with both numerical integration of continuous-time stochastic differential equations and discrete-generation Wright-Fisher simulations.
Finally, we use the multi-allele solution to predict the fate of a novel beneficial mutation in the clonal interference scenario -- competition with other mutations that are simultaneously present in the evolving population.



\section{Mathematical background}

Our presentation in this section closely follows previous treatments~\citep{Chernyak2006,Mustonen2010} and is reproduced here for completeness.
Consider evolution of a population with $M$ possible alleles or genotypes, each one represented at a frequency $x^i \ge 0$ in the population, such that $\sum_{i=1}^M x^i = 1$ and the genotype frequencies are confined to the $M-1$ dimensional simplex $\Sigma_{M-1}$. Because of the normalization condition, only $M-1$ frequencies are linearly independent, chosen here to be $\mathbf{x} = (x^1, \dots, x^{M-1})$ without loss of generality.
We will use the Einstein summation convention with upper and lower indices, reserving Greek letters $\alpha, \beta, \gamma, \dots$ for sums over $M-1$ linearly independent genotypes:
$F^\alpha G_\alpha = \sum_{\alpha=1}^{M-1} F^\alpha G_\alpha$ and Latin letters
$i, j, k, \dots$ for sums over all $M$ genotypes: $F^i G_i = \sum_{i=1}^M F^i G_i$ (in both cases, $\mathbf{F}$ and $\mathbf{G}$ are arbitrary vector-valued functions of the genotype frequencies $\mathbf{x}$ and time $t$).
The sums will be written out explicitly in all cases where ambiguity arises.

\subsection{Deterministic evolutionary dynamics: general case}

The evolution of genotype frequencies under selection and mutations and in the absence of genetic drift (stochastic effects due to the finite population size) is described
by the replicator-mutator equations~\citep{Page2002,Nowak2006book}:
\begin{equation} \label{repl:mut:eq:new}
\dot{x}^i = \sum_{j=1}^M x^j f_j Q^{j}_{~i} - \langle f \rangle x^i,
\end{equation}
where $\dot{x}^i = {d x^i}/{dt}$ is the time derivative of the genotype frequency $x^i$, $f_i = f_i (\mathbf{x}, t)$ is the fitness value of genotype $i$, and $\langle f \rangle = x^j f_j$ is the average fitness of the population.
The probability per unit time that genotype $i$ gives rise to genotype $j$ due to mutation is given by $Q^{i}_{~j}$, the $(i,j)$ element of the $M \times M$ mutation matrix $Q$. Note that each
row of the mutation matrix sums to one: $\sum_{i=1}^M Q^{j}_{~i} = 1, \forall j$, such that Eq.~\eqref{repl:mut:eq:new} preserves the normalization of genotype frequencies: $\frac{d}{dt} \sum_{i=1}^M x^i = 0$.
Here, we shall focus on the symmetric mutation rates between genotypes $i$ and $j$: $Q^{j}_{~i} = Q^{i}_{~j}$.
Note that all sums are written explicitly in this section.

Moreover, we assume that each genotype mutates into every one of its $N_b$ mutational neighbors with the same rate $\mu$:
\begin{equation} \label{Q:1:new}
Q = 
\begin{pmatrix}
1 - N_b \mu & \mu & 0 & \dots & \mu & \dots \\
\mu & 1 - N_b \mu & \mu & \dots \\
0 & \mu & 1 - N_b \mu & \dots \\
\dots & & & \\
\mu & \dots &  & \\
\dots & & &
\end{pmatrix},
\end{equation}
where each row contains $N_b$ non-zero off-diagonal elements of the magnitude $\mu$. With this mutation matrix, the replicator-mutator equations become
\begin{equation} \label{repl:mut:eq:1:new}
\dot{x}^i = \mu N_b \left[ \langle f \rangle_i - x^i f_i \right] + x^i \left[ f_i - \langle f \rangle \right],
\end{equation}
where $\langle f \rangle_i = N_b^{-1} \sum_{j \in \text{nnb} (i)} x^j f_j$ is the frequency-weighted average over all mutational neighbors of genotype $i$ and $\text{nnb} (i)$ denotes its
set of mutational neighbors of size $N_b$. The first term on the right-hand side of Eq.~\eqref{repl:mut:eq:1:new} is the contribution due to mutations. Note that if $\mu \ll 1$ and the selection
is weak: $f_M = 1$ and $f_\alpha = 1 + s_\alpha$  with $|s_\alpha| \ll 1$ $(\forall \alpha)$, we can neglect the $\mathcal{O} (\mu s_\alpha)$ terms in Eq.~\eqref{repl:mut:eq:1:new}, yielding
\begin{equation} \label{repl:mut:eq:2:new}
\dot{x}^i \simeq \mu~[ \!\!\!\sum_{j \in \text{nnb} (i)} x^j - N_b x^i ] + x^i \left[ f_i - \langle f \rangle \right].
\end{equation}
In this limit, mutation and selection forces become additive. 

\subsection{Deterministic evolutionary dynamics under selection only}

The evolution of genotype frequencies under selection (i.e., in the absence of mutations and genetic drift)
is described by a system of replicator equations~\citep{Eigen1988,Eigen1989,Hofbauer1998,Page2002,Nowak2006book},
which follows from Eq.~\eqref{repl:mut:eq:1:new} by setting $\mu = 0$:
\begin{equation} \label{repl:eq:1}
\dot{x}^i = x^i \left[ f_i - \langle f \rangle \right].
\end{equation}
Here, we focus on the case where the fitness functions are independent of both the genotype frequencies $\mathbf{x}$ and the time $t$.

For linearly independent frequencies, Eq.~\eqref{repl:eq:1} transforms into
\begin{equation} \label{repl:eq:2}
\dot{x}^\alpha = g^{\alpha \beta} s_\beta = s^\alpha,
\end{equation}
where $s_\beta = f_\beta - f_M$ are the selection coefficients and
\begin{equation} \label{metric:inv}
g^{\alpha \beta} (\mathbf{x}) = \left\{
\begin{array}{cc}
-x^\alpha x^\beta & \text{ if } \alpha \ne \beta,  \\
x^\alpha (1 - x^\alpha) & \text{ if } \alpha = \beta.
\end{array}
\right.
\end{equation}

The $g^{\alpha \beta}$ coefficients in Eq.~\eqref{repl:eq:2} form an $(M-1) \times (M-1)$ matrix $g$ whose inverse, $g^{-1}$, is a metric tensor on the simplex $\Sigma_{M-1}$~\citep{Hofbauer1998}.
Note that both $g$ and $g^{-1}$ are symmetric matrices.
Denoting the matrix elements of $g^{-1}$ as $g_{\alpha \beta}$ (such that $g^{\alpha \beta} g_{\beta \gamma} = \delta^{\alpha}_{~\gamma}$, where $\delta^{\alpha}_{~\gamma}$ are the elements of a unit matrix), we observe that the metric tensor and its inverse can be used to raise or lower tensor indices: $D^{\alpha} = g^{\alpha \beta} D_\beta$, $D_{\alpha} = g_{\alpha \beta} D^\beta$, where $D^{\alpha}$ and $D_{\alpha}$
are the components of an arbitrary contravariant vector and its covariant counterpart, respectively. In particular, $g^{\alpha \beta} s_\beta = s^\alpha$ in Eq.~\eqref{repl:eq:2} are the contravariant selection coefficients.

Next, we observe that since the covariant selection coefficients $s_\alpha$ are independent of the genotype frequencies $\mathbf{x}$ and time $t$,
they can always be represented as gradients of the fitness potential $F(\mathbf{x})$:
\begin{equation} \label{F:pot:1}
s_\alpha = \partial_\alpha F(\mathbf{x}),
\end{equation}
where $F(\mathbf{x}) = s_\gamma x^\gamma$ and $\partial_\alpha = {\partial}/{\partial x^\alpha}$ are covariant partial derivatives with respect to the genotype frequencies.

\subsection{Deterministic evolutionary dynamics under mutation only}

According to Eq.~\eqref{repl:mut:eq:2:new}, the evolution of genotype frequencies under mutations only (i.e., in the absence of selection and genetic drift) is described by
\begin{equation} \label{repl:mut:eq:0}
\dot{x}^i = \mu~[ \!\!\!\sum_{j \in \text{nnb} (i)} x^j - N_b x^i ],
\end{equation}
where the first term on the right-hand side is the gain term due to mutations of all other genotypes into genotype $i$ and the second term is the loss term due to mutations of genotype $i$ into all other genotypes.
For simplicity, we shall assume in the following that the mutation rate $\mu$ is the same between any two genotypes (i.e., $N_b = M-1$), resulting in
\begin{equation} \label{mut:eq:1}
\dot{x}^i = \mu \sum_{\substack{j=1 \\ j \ne i}}^M x^j - \mu (M-1) x^i,
\end{equation}

In terms of linearly independent frequencies, we obtain:
\begin{equation} \label{mut:eq:2}
\dot{x}^\alpha = \mu - \mu M x^\alpha \equiv m^\alpha (\mathbf{x}).
\end{equation}

Note that $m^\alpha = g^{\alpha \beta} m_\beta$, where $m_\beta (\mathbf{x}) = \partial_\beta \mathsf{M} (\mathbf{x})$. In the last expression, 
$\mathsf{M} (\mathbf{x})$ is the mutation potential given by~\citep{Baxter2007,Mustonen2010}:
\begin{equation} \label{mut:pot}
\mathsf{M} (\mathbf{x}) = \mu \sum_{i=1}^M \log(x^i).
\end{equation}

\subsection{Stochastic evolutionary dynamics under selection and mutation}
\noindent
\textbf{The Fokker-Planck equation.}  The evolution of finite-size populations is described by a time-dependent probability distribution of genotype frequencies, $P(\mathbf{x},t)$, which obeys the Fokker-Planck
(forward Kolmogorov) equation
of the Kimura-Ohta type~\citep{Kimura1971,Kimura1983,Mustonen2010}:
\begin{equation} \label{FP:eq:1}
\frac{\partial P(\mathbf{x},t)}{\partial t} = \partial_\alpha \left[ \frac{1}{2N} \partial_\beta g^{\alpha \beta} (\mathbf{x}) - v^\alpha (\mathbf{x}) \right] P(\mathbf{x},t),
\end{equation}
where $N$ is the population size and $v^\alpha (\mathbf{x}) = s^\alpha (\mathbf{x}) + m^\alpha (\mathbf{x})$ is the sum of contravariant selection and mutation coefficients given by Eqs.~\eqref{repl:eq:2} and \eqref{mut:eq:2}, respectively.
Note that the partial derivatives in Eq.~\eqref{FP:eq:1} act on everything to their right, e.g. $\partial_\alpha \partial_\beta g^{\alpha \beta} P = \partial_\alpha (\partial_\beta (g^{\alpha \beta} P))$, etc.

Eq.~\eqref{FP:eq:1} is consistent with the following Langevin equation, understood in the It\^{o} sense: 
\begin{equation} \label{Lng:eq:1}
\dot{x}^\alpha = v^\alpha + \xi^\alpha.
\end{equation}
Here, $\xi^\alpha$ is $\delta$-correlated noise with zero mean and the pair correlation function given by
\begin{equation} \label{Lng:pair}
\langle \xi^\alpha (t) \xi^\beta (t') \rangle = \frac{g^{\alpha \beta}}{N} \delta (t-t'),
\end{equation}
where $\langle \dots \rangle$ denotes the average over independent noise realizations.

We can use a standard linear algebra relation~\citep{bishop2006pattern} $\partial_\beta \Omega_g = \text{Tr} \left( g^{-1} \partial_\beta g \right)$, where $\Omega_g = \log \det (g)$ and $\text{Tr} (\cdot)$ denotes a matrix trace,
to obtain the following identity:
\begin{equation} \label{Omega:id}
g^{\alpha \beta} (\partial_\beta \Omega_g) = \partial_\beta g^{\alpha \beta}.
\end{equation}

Eq.~\eqref{Omega:id} allows us to re-write Eq.~\eqref{FP:eq:1} as
\begin{equation} \label{FP:eq:2}
\frac{\partial P(\mathbf{x},t)}{\partial t} = \partial_\alpha \left[ \frac{1}{2N} g^{\alpha \beta} (\mathbf{x}) \partial_\beta  - \widetilde{v}^\alpha (\mathbf{x}) \right] P(\mathbf{x},t),
\end{equation}
where
\begin{equation} \label{v:tilde}
\widetilde{v}^\alpha = v^\alpha - \frac{1}{2N} g^{\alpha \beta} (\partial_\beta \Omega_g)
\end{equation}
is the effective force due to selection, mutation, and genetic drift. Eq.~\eqref{FP:eq:2} can be rewritten as the probability conservation law:
\begin{equation} \label{FP:eq:3}
\frac{\partial P}{\partial t} + \partial_\alpha J^\alpha = 0,
\end{equation}
where
\begin{equation} \label{J:cur}
J^\alpha = \tilde{v}^\alpha P - \frac{1}{2N} g^{\alpha \beta} (\partial_\beta P)
\end{equation}
is the probability current. \\

\noindent
\textbf{The two-allele case.} Our mathematical framework yields well-known results~\citep{Kimura1955,Kimura1962,Kimura1964} in the case of two genotypes or alleles ($M=2$),
the mutant allele $A_1$ with frequency $x$ and fitness $f_1 = 1 + s$ ($s$ is the constant selection coefficient) and the wild-type allele $A_2$ with frequency $1 - x$ and fitness $f_2 = 1$.
The two genotypes mutate into one another with the mutation rate $\mu$. In this case, the standard formulation of the one-dimensional Fokker-Planck equation for the evolution
of gene frequencies is given by~\citep{Kimura1955,Kimura1964}:
\begin{equation} \label{FP:Kimura}
\frac{\partial P}{\partial t} + \frac{\partial J}{\partial x} = 0,
\end{equation}
where
\begin{equation} \label{J:Kimura}
J(x,t) = \mathcal{M} (x) P(x,t) - \frac{1}{2} \frac{\partial}{\partial x} [\mathcal{V} (x) P(x,t)].
\end{equation}
Here, $\mathcal{M} (x) = sx(1-x) + \mu(1-2x)$ is the mean change in gene frequency $x$ per generation due to mutation and selection, and
$\mathcal{V} (x) = {x(1-x)}/{N}$ is the variance of the change in gene frequency.
Note that $x^1 \to x$, $\mathbf{x} \to x$ and $J^1 (\mathbf{x},t) \to J(x,t)$ in this notation.
In this case, the steady-state probability distribution is the equilibrium distribution $P_\text{eq} (x)$ characterized by the zero probability current: $J_\text{eq} (x) = 0$. 
The zero-current condition yields a first-order ordinary differential equation for $P_\text{eq} (x)$, which can be solved to yield
\begin{equation} \label{Peq:Kimura}
P_\text{eq} (x) = \frac{1}{Z} [x(1-x)]^{2 N \mu - 1} e^{2Nsx},
\end{equation}
where $Z$ is the normalization constant. In the $s=0$, $2 N \mu \ll 1$ limit, $P_\text{eq} (x)$ is given by a well-known U-shaped distribution~\citep{Kimura1964,Kimura1983}:
\begin{equation} \label{Peq:Kimura:U}
P_\text{eq}^0 (x) = \frac{1}{Z} \frac{1}{x(1-x)}.
\end{equation}

On the other hand, Eq.~\eqref{J:cur} becomes:
\begin{equation} \label{J:cur:eq:1}
J^{1}_\text{eq} = [ s^1 + m^1 - \frac{1}{2N} g^{11} (\partial \Omega_g) ] P_\text{eq} - \frac{1}{2N} g^{11} (\partial P_\text{eq}) = 0,
\end{equation}
where $\partial = \partial/\partial x$. According to Eq.~\eqref{metric:inv}, $g^{11} (x) = x(1 - x)$, yielding $\Omega_g (x) = \log [x (1-x)]$. Furthermore, $s_1 = f_1 - f_2 = s$, 
resulting in $F(x) = s x$ (Eq.~\eqref{F:pot:1}). Finally, the contravariant selection coefficient $s^1 = g^{11} s_1 = s x(1-x)$.
The contravariant mutation force is given by $m^1 = g^{11} \partial \mathsf{M} (x) = \mu (1 - 2x)$, with the mutation potential $\mathsf{M} (x) = \mu \log[x (1-x)]$ (Eq.~\eqref{mut:pot}).

Noting that $g^{11} \partial \Omega_g = \partial g^{11}$ (Eq.~\eqref{Omega:id}), we can rewrite Eq.~\eqref{J:cur:eq:1} as
\begin{equation} \label{J:cur:eq:2}
[s x (1-x) + \mu (1-2x)] P_\text{eq} - \frac{1}{2} \partial \left( \frac{g^{11}}{N} P_\text{eq} \right) = 0,
\end{equation}
which is identical to Eq.~\eqref{J:Kimura} and is therefore solved by $P_\text{eq} (x)$ in Eq.~\eqref{Peq:Kimura}. A key observation is that the equilibrium probability distribution
can be written as a Boltzmann-like distribution involving the fitness potential $F(x)$, the mutation potential $\mathsf{M} (x)$, and the genetic drift potential $\Omega_g (x)$~\citep{Mustonen2010}:
\begin{equation} \label{Peq:pot}
P_\text{eq} (x) = \frac{1}{Z} e^{2 N \mathsf{M} (x) - \Omega_g(x) + 2 N F(x)}.
\end{equation}
Eq.~\eqref{Peq:pot} satisfies Eq.~\eqref{J:cur:eq:1} and is equivalent to Eq.~\eqref{Peq:Kimura}, as can be verified by direct substitution. \\

\noindent
\textbf{The steady-state (equilibrium) distribution.} Let us now assume that
\begin{equation} \label{P:ss}
P_\text{ss} (\mathbf{x}) = e^{-\varphi(\mathbf{x})}
\end{equation}
at steady state. Eq.~\eqref{FP:eq:2} can be re-written as:
\begin{equation} \label{FP:eq:4}
\frac{\partial P}{\partial t} = -\frac{1}{2} \partial_\alpha \left[ g^{\alpha \beta} (2 \widetilde{v}_\beta - \frac{1}{N} \partial_\beta)  P \right].
\end{equation}

At steady state,
\begin{equation} \label{J:ss}
J^\alpha_\text{ss} = \frac{1}{2} g^{\alpha \beta} (2 \widetilde{v}_\beta - \frac{1}{N} \partial_\beta) e^{-\varphi} = \left[ \widetilde{v}^\alpha + \frac{1}{2N} g^{\alpha \beta} (\partial_\beta \varphi) \right] e^{-\varphi}.
\end{equation}
If all the forces in Eq.~\eqref{FP:eq:4} can be represented as gradients of the corresponding potentials, the resulting steady-state probability distribution $P_\text{eq} (\mathbf{x})$
is Boltzmann-like~\citep{Mustonen2010} and is characterized by detailed balance and zero probability currents. Indeed,
since $\widetilde{v}_\alpha = - \frac{1}{2} \partial_\alpha U$, where $U = -2 (F + \mathsf{M}) + \Omega_g/N$, Eq.~\eqref{FP:eq:4} has a steady-state solution if
$\varphi = N U - C$, where $C = - \log Z$ is the normalization constant. Moreover, $J^\alpha_\text{ss} = 0$, indicating that
the steady-state distribution is at equilibrium.

Thus,
\begin{equation} \label{Peq:pot:gen}
P_\text{eq} (\mathbf{x}) = \frac{1}{Z} e^{2 N \mathsf{M} (\mathbf{x}) - \Omega_g(\mathbf{x}) + 2 N F(\mathbf{x})}
\end{equation}
is a generalization of Eq.~\eqref{Peq:pot} to an arbitrary number of alleles or genotypes. \\


\noindent
\textbf{Fixation probabilities: two alleles.} Eq.~\eqref{FP:Kimura} corresponds to the backward Kolmogorov equation~\citep{Gardiner2009}, which can be used to derive an ordinary differential equation for $P_\text{fix}$, the fixation
probability of the mutant allele $A_1$~\citep{Weiss1966,Gillespie2004,Crow2009}:
\begin{equation} \label{FP:fixprob}
\mathcal{M} (p) \frac{d P_\text{fix} (p)}{dp} + \frac{1}{2} \mathcal{V} (p) \frac{d^2 P_\text{fix} (p)}{dp^2} = 0,
\end{equation}
where $p$ is the initial ($t=0$) frequency of the $A_1$ allele in a population of $N$ individuals. In the absence of mutations, $\mathcal{M} (p) = sp(1-p)$ and $\mathcal{V} (p) = {p(1-p)}/{N}$, so that
Eq.~\eqref{FP:fixprob} is solved by
\begin{equation} \label{Kimura:fixprob}
P_\text{fix} (p) = \frac{1 - e^{-2Nsp}}{1 - e^{-2Ns}}.
\end{equation}
This is the celebrated Kimura's formula for the fixation probability in a two-allele system~\citep{Kimura1962}. The analytical solution for the fixation probabilities of more than two alleles
has not yet been obtained~\citep{Crow2009}, despite an early claim to the contrary~\citep{Kimura1964}.

\subsection{Path integral approach to the Fokker-Planck equation}

The Fokker-Planck equation (Eq.~\eqref{FP:eq:2}) admits a path integral solution~\citep{Chernyak2006,Mustonen2010}. Consider
stochastic trajectories that describe the evolution of the population from the initial time $-\tau$ to the final time $\tau$. Each trajectory is characterized by a set of genotype frequencies
observed over subsequent generations separated by a time interval $\Delta t$: $X = \{ \mathbf{x}_t \}_{-\tau}^{\tau}$, where $t = \{ -\tau, -\tau+\Delta t, \dots , \tau-\Delta t, \tau \}$. In the $\Delta t \to 0$ limit,
the probability of observing a specific trajectory starting from a fixed initial state $\mathbf{x}_{-\tau}$ is given by~\citep{Chernyak2006}:
\begin{equation} \label{Path:eq:1}
P^F [X | \mathbf{x}_{-\tau}] = \mathcal{N} e^{-N \int_{-\tau}^{\tau} dt S_F (\mathbf{x}_t, \dot{\mathbf{x}}_t)},
\end{equation}
where $\mathcal{N}$ is the normalization constant which ensures correct normalization over the ensemble of all paths
($\int \mathcal{D} X P^F [X | \mathbf{x}_{-\tau}] = 1$), $\dot{\mathbf{x}}_t = {d \mathbf{x}_t}/{dt}$ is the time derivative of the vector of genotype frequencies,
and $S_F$ is the `action' of the path propagating forward in time. The integral is interpreted as the limit of a discrete sum written in the mid-point (Stratonovich) discretization~\citep{Gardiner2009},
which yields~\citep{Chernyak2006}:
\begin{equation} \label{SF:eq:1}
S_F = \frac{1}{2} (\dot{x}^\alpha - \widetilde{v}^\alpha) g_{\alpha \beta} (\dot{x}^\beta - \widetilde{v}^\beta) +  \frac{1}{2} \partial_\alpha \widetilde{v}^\alpha.
\end{equation}

If the initial state $\mathbf{x}_{-\tau}$ is sampled from an arbitrary distribution $Q (\mathbf{x}_{-\tau}) = e^{-\psi (\mathbf{x}_{-\tau})}$~\citep{Seifert2005,Seifert2012}, we can write the total probability of observing the path $X$ as
\begin{equation} \label{Ptot:eq}
P^F [X] = P^F [X | \mathbf{x}_{-\tau}] Q (\mathbf{x}_{-\tau}).
\end{equation}

Now, consider the reverse path $X^+ = \{ \mathbf{x}^+_t \}_{-\tau}^{\tau}$, where $\mathbf{x}^+_t = \mathbf{x}_{-t}$. Then the probability of the reverse path conditioned on the initial state at time $\tau$,
$\mathbf{x}^+_{-\tau} = \mathbf{x}_{\tau}$, is given by
\begin{eqnarray} \label{Path:eq:2}
P^R [X^+ | \mathbf{x}^+_{-\tau}] &=& \mathcal{N} e^{-N \int_{-\tau}^{\tau} dt S_F (\mathbf{x}^+_t, \dot{\mathbf{x}}^+_t)} \\
&=& \mathcal{N} e^{-N \int_{-\tau}^{\tau} dt S_R (\mathbf{x}_t, \dot{\mathbf{x}}_t)}, \nonumber
\end{eqnarray}
where $S_R (\mathbf{x}_t, \dot{\mathbf{x}}_t) \equiv S_F (\mathbf{x}_t, -\dot{\mathbf{x}}_t)$ is the `action' of the path propagating backward in time.
Recalling that $\widetilde{v}^\alpha =  \widetilde{v}^\alpha (\mathbf{x})$ (i.e., it does not depend on $\dot{\mathbf{x}}$), we obtain:
\begin{equation} \label{SF:SR:diff:eq}
S_F (\mathbf{x}, \dot{\mathbf{x}}) - S_R (\mathbf{x}, \dot{\mathbf{x}}) = -2 \dot{x}^\alpha \widetilde{v}_\alpha.
\end{equation}

Thus, the ratio of the conditional probabilities for the forward and reverse paths adopts a particularly simple form:
\begin{equation} \label{PF:PR:cond:eq}
\frac{P^F [X | \mathbf{x}_{-\tau}]}{P^R [X^+ | \mathbf{x}^+_{-\tau}]} = e^{2 N  \int_{-\tau}^{\tau} dt \dot{x}^\alpha \widetilde{v}_\alpha}.
\end{equation}

Furthermore, if the final state $\mathbf{x}_{\tau}$ is sampled from the same distribution $Q (\mathbf{x}_{\tau})$ as the initial state,
\begin{equation} \label{PF:PR:tot:eq}
\frac{P^F [X]}{P^R [X^+]} = \frac{Q (\mathbf{x}_{-\tau})}{Q (\mathbf{x}_{\tau})} e^{2 N  \int_{-\tau}^{\tau} dt \dot{x}^\alpha \widetilde{v}_\alpha}. 
\end{equation}
In particular, if $Q = P_\text{eq}$, we obtain the detailed balance condition, $P^F [X] = P^R [X^+]$. Indeed, let us compute
\begin{equation} \label{aux:1}
2 N \int_{-\tau}^{\tau} dt \dot{x}^\alpha \widetilde{v}_\alpha = 2 N \int_F d x^\alpha \widetilde{v}_\alpha,
\end{equation}
where $\int_F$ denotes an integral over the forward path. This integral should be understood as the limiting value of the sum over $K$ time intervals of length $\Delta$:
\begin{equation} \label{aux:2}
\int_F d x^\alpha \widetilde{v}_\alpha = \lim_{\substack{\Delta \to 0, \\ K \to \infty}} ~\sum_{i=0}^{K-1} \Delta x^\alpha_i \widetilde{v}_\alpha (\widetilde{\mathbf{x}}_i),
\end{equation}
where $\Delta x^\alpha_i = x^\alpha_{i+1} - x^\alpha_{i}$ and $\widetilde{x}^\alpha_i = \frac{1}{2} (x^\alpha_{i} + x^\alpha_{i+1})$. In this notation,
$\widetilde{\mathbf{x}}_i = (\widetilde{x}^{1}_{i}, \dots , \widetilde{x}^{M-1}_{i})$ is a vector of linearly independent genotype frequencies at midpoints along the path.
Since $\widetilde{v}_\alpha$ is a gradient of the total potential in this case,
\begin{eqnarray} \label{aux:3}
2 N \int_F d x^\alpha \widetilde{v}_\alpha &=& 2 N [\mathsf{M} (\mathbf{x}_{\tau}) - \mathsf{M} (\mathbf{x}_{-\tau}) + F(\mathbf{x}_{\tau}) - F(\mathbf{x}_{-\tau})] \nonumber \\
&-& [\Omega_g(\mathbf{x}_{\tau}) - \Omega_g(\mathbf{x}_{-\tau})].
\end{eqnarray}
The exponentiated right-hand side in Eq.~\eqref{aux:3} cancels with the ${P_\text{eq} (\mathbf{x}_{-\tau})} / {P_\text{eq} (\mathbf{x}_{\tau})}$ prefactor in Eq.~\eqref{PF:PR:tot:eq} (cf. Eq.~\eqref{Peq:pot:gen}),
such that the detailed balance condition is recovered.

In general,
\begin{equation} \label{PF:PR:tot:eq:gen}
\frac{P^F [X]}{P^R [X^+]} = e^{N  \int_{-\tau}^{\tau} dt \dot{x}^\alpha A_\alpha},
\end{equation}
where we defined $A_\alpha = 2 \widetilde{v}_\alpha + \frac{1}{N} \partial_\alpha \psi$ and used the fact that
\begin{equation} \label{delta:phi:eq}
\Delta \psi \equiv \psi (\mathbf{x}_{\tau}) - \psi (\mathbf{x}_{-\tau}) = \int_{-\tau}^{\tau} dt \dot{x}^\alpha \partial_\alpha \psi.
\end{equation}
Note that $A_\alpha = 0$ in the $Q = P_\text{eq}$ case, consistent with detailed balance.

Let us define an integrated quantity that characterizes deviations from equilibrium over a forward path~\citep{Chernyak2006}:
\begin{equation} \label{RF:eq}
R^F (X) = \int_{-\tau}^{\tau} dt \dot{x}^\alpha A_\alpha = \int_F d x^\alpha A_\alpha.
\end{equation}
Then Eq.~\eqref{PF:PR:tot:eq:gen} can be written as
\begin{equation} \label{PF:PR:tot:eq:short}
\frac{P^F [X]}{P^R [X^+]} = e^{N R^F (X)}.
\end{equation}
For the reverse path, we have
\begin{equation} \label{RR:eq}
R^R (X^+) = \int_{\tau}^{-\tau} dt \dot{x}^\alpha A_\alpha = -R^F (X),
\end{equation}
as can be seen with the help of Eq.~\eqref{aux:2}.
The probability distribution of $R$ over the ensemble of all forward paths is given by
\begin{eqnarray} \label{rhoF:eq}
\rho^F (R) &=& \int \mathcal{D} X P^F [X] \delta (R - R^F (X)) \\
&=& e^{NR} \int \mathcal{D} X^+ P^R [X^+] \delta (R + R^R (X^+)) \nonumber \\
&=& e^{NR} \rho^R (-R), \nonumber
\end{eqnarray}
where $\rho^R (\widetilde{R})$ is the distribution of $\widetilde{R}$ over the ensemble of reverse paths; here, we have made use of Eqs.~\eqref{PF:PR:tot:eq:short} and \eqref{RR:eq} and the fact that
$\mathcal{D} X = \mathcal{D} X^+$. Thus, the distributions of $R$ over forward paths and $-R$ over reverse paths 
are simply related:
\begin{equation} \label{rhoF:rhoR:eq}
\frac{\rho^F (R)}{\rho^R (-R)} = e^{NR}.
\end{equation}
This is the fluctuation theorem, originally derived for physical systems characterized by non-equilibrium Langevin dynamics~\citep{Crooks1999,Crooks2000,Seifert2005,Chernyak2006}.
Multiplying both sides of Eq.~\eqref{rhoF:rhoR:eq} by ${\rho^R (-R)} e^{-NR}$ and integrating over all values of $R$, we obtain the integrated fluctuation theorem~\citep{Jarzynski1997,Seifert2005,Chernyak2006,Pitaevskii2011}:
\begin{equation} \label{Fluct:int}
\int_{-\infty}^{\infty} dR \rho^F (R) e^{-NR} \equiv \langle e^{-NR} \rangle_F = 1.
\end{equation}

Note that, according to Eq.~\eqref{Fluct:int}, $\rho^F (R) e^{-NR}$ is a normalized probability distribution which is in fact equal to $\rho^R (-R)$, the probability of observing a value of $-R$ in the ensemble of reverse paths
(Eq.~\eqref{rhoF:rhoR:eq}). Furthermore, although Eq.~\eqref{Fluct:int} was derived using both forward and reverse paths, it is a constraint imposed solely on the ensemble of forward paths. 

Applying Jensen's inequality, $\langle e^y \rangle \ge e^{\langle y \rangle}$, to the rightmost relation in Eq.~\eqref{Fluct:int}, we obtain:
\begin{equation} \label{Fluct:int:J}
\langle R \rangle_F \ge 0.
\end{equation}
This is the equivalent of the second law of thermodynamics~\citep{Jarzynski1997,Mustonen2010,Pitaevskii2011}.

In our case, all the forces are conservative, resulting in
\begin{eqnarray} \label{R:comp}
R^F (X) &=& 2 [ \mathsf{M} (\mathbf{x}_{\tau}) - \mathsf{M} (\mathbf{x}_{-\tau}) ] + 2 [ F(\mathbf{x}_{\tau}) - F(\mathbf{x}_{-\tau}) ] \nonumber \\
              &-& \frac{1}{N} [ \Omega_g (\mathbf{x}_{\tau}) - \Omega_g (\mathbf{x}_{-\tau})] + \frac{1}{N} \Delta \psi.
\end{eqnarray}
Thus, the contribution of each path is fully defined by its initial and final points. 

\section{Results}

\subsection{Fluctuation theorem constraint on fixation probabilities}
We now consider the two-allele system described by Eq.~\eqref{FP:Kimura}. In the absence of mutations ($\mu = 0$), all paths starting from the initial frequency $p$ of the mutant allele $A_1$ in a population
of $N$ individuals result either in fixation of $A_1$ ($x_f = 1$) or its extinction ($x_f = 0$). Then
\begin{equation} \label{rho:F}
\rho^F (R) = P_0 \delta (R - R^{F} (x_f = 0)) + P_1 \delta (R - R^{F} (x_f = 1)),
\end{equation}
where $P_0$ and $P_1$ correspond to the fraction of all paths that end up in the $x_f = 0$ and $x_f = 1$ states, respectively. Since $\rho^F (R)$ is normalized,
$\int_{-\infty}^{\infty} dR \rho^F (R) = 1$, we obtain $P_0 + P_1 = 1$.

According to Eq.~\eqref{R:comp} with $\mu = 0$,
\begin{eqnarray} \label{R:comp:fix}
R^{F} (x_f) &=& 2 s (x_f - p) - \frac{1}{N} \log \frac{x_f (1 - x_f)}{p (1 - p)} \nonumber \\
&+& \frac{1}{N} \log \frac{Q(p)}{Q(x_f)}.
\end{eqnarray}
Now, if we choose $Q(x)$ to be $P_\text{eq} (x)$ \emph{in the absence of selection and with vanishing mutation rates}, Eq.~\eqref{Peq:Kimura:U}, we obtain:
\begin{equation} \label{R:comp:aux1}
\log \frac{Q(p)}{Q(x_f)} = \log \frac{x_f (1 - x_f)}{p (1 - p)},
\end{equation}
such that the last two terms in Eq.~\eqref{R:comp:fix} cancel each other and $e^{-N R^{F} (x_f)} = e^{-2Ns (x_f - p)}$. As a result, Eq.~\eqref{Fluct:int} becomes
\begin{equation} \label{Eq:fluxes}
P_1 e^{-2Ns (1-p)} + (1 - P_1) e^{2Ns p} = 1,
\end{equation}
which immediately yields
\begin{equation} \label{Eq:fluxes:final}
P_1 = \frac{1 - e^{-2Nsp}}{1 - e^{-2Ns}}.
\end{equation}
Eq.~\eqref{Eq:fluxes:final} is the same as Kimura's formula for the fixation probability in a two-allele system~\citep{Kimura1962}, Eq.~\eqref{Kimura:fixprob},
with $P_1 = P_\text{fix} (p)$, the fixation probability of the mutant allele. Note that $\lim_{s \to 0} P_1 = p$, a well-known result that the fixation probability
of a neutral mutant is equal to its initial frequency in the population~\citep{Gillespie2004}.

For a system with an arbitrary number $M$ of alleles, we set $f_M = 1$ for the wild-type allele $A_M$ and $f_\alpha = 1 + s_\alpha~(\alpha = 1 \dots M-1)$ with $|s_\alpha| \ll 1$
for $M-1$ mutant alleles $A_\alpha$. Then $F(\mathbf{x}_{-\tau}) = \sum_{\alpha=1}^{M-1} s_\alpha x^{\alpha}_{0}$ and $F(\mathbf{x}_{\tau}) = f_j - 1$, where $x^{\alpha}_{0}$ is the initial frequency of the mutant
allele $A_\alpha$ at time $-\tau$ and $j = 1 \dots M$ is the index of the allele fixed at time $\tau$.
Using
\begin{equation} \label{Peq:nomut}
Q (\mathbf{x}) = P_\text{eq}^0 (\mathbf{x}) = \frac{1}{Z} e^{-\Omega_g (\mathbf{x})}
\end{equation}
(cf. Eq.~\eqref{Peq:pot:gen}), we obtain from Eq.~\eqref{Fluct:int}:
\begin{equation} \label{Eq:fluxes:final:M}
\sum_{j} P_{\text{fix}}^{j} e^{-2 N f_j} = e^{-2N (1 + \sum_{\alpha} s_\alpha x^{\alpha}_{0})},
\end{equation}
where $P_{\text{fix}}^{j}$ is the fixation probability of the allele $A_j$:
\begin{equation} \label{Eq:Pfix:norm}
\sum_{j=1}^M P_{\text{fix}}^{j} = 1.
\end{equation}

Eqs.~\eqref{Eq:fluxes:final} and \eqref{Eq:fluxes:final:M} constitute the first major theoretical result of this work. Eq.~\eqref{Eq:fluxes:final} shows that the integrated fluctuation theorem, together with the normalization condition,
can be used to obtain Kimura's formula without solving the backward Kolmogorov equation. Eq.~\eqref{Eq:fluxes:final:M} provides a constraint which must be obeyed by any valid set of fixation probabilities.

Eq.~\eqref{Eq:fluxes:final:M} can be re-written as:
\begin{equation} \label{Eq:fluxes:final:M2}
\mathcal{F} (\{ P_{\text{fix}}^{\alpha} \}, \{ s_\alpha \}, \{ x^{\alpha}_{0} \}) \equiv
e^{-2N \sum_{\alpha} s_\alpha x^{\alpha}_{0}} - \sum_{\alpha} \left( e^{-2 N s_\alpha} -1 \right) P_{\text{fix}}^{\alpha} = 1.
\end{equation}
In the weak-selection limit, $N |s_\alpha| \ll 1$, $\forall \alpha$, Eq.~\eqref{Eq:fluxes:final:M2} simplifies to:
\begin{equation} \label{Eq:fluxes:final:M3}
\sum_{\alpha} s_\alpha P_{\text{fix}}^{\alpha} \simeq \sum_{\alpha} s_{\alpha} x^{\alpha}_{0}.
\end{equation}

Furthermore, if all the alleles have the same selection coefficient, $s_\alpha = s$, including the strictly neutral case of $s \to 0$, Eq.~\eqref{Eq:fluxes:final:M3} yields:
\begin{equation} \label{Eq:fluxes:final:M4}
\sum_{\alpha} P_{\text{fix}}^{\alpha} \simeq \sum_{\alpha} x^{\alpha}_{0},
\end{equation}
which is trivially solved by $P_{\text{fix}}^{\alpha} = x^{\alpha}_{0}$ due to the symmetry between all the mutant alleles. 
Note however that in general, Eq.~\eqref{Eq:fluxes:final:M3} alone cannot be used to determine the full set of fixation probabilities.

In the case where the initial frequencies of all the mutant alleles are equivalent, $x^{\alpha}_{0} = x_{0}$, Eq.~\eqref{Eq:fluxes:final:M3} takes the form of a weighted average:
\begin{equation} \label{Eq:fluxes:final:M5}
\frac{\sum_{\alpha} s_\alpha P_{\text{fix}}^{\alpha}}{\sum_{\alpha} s_\alpha} \simeq x_{0}.
\end{equation}

Finally, in the case when one of the mutants is much more beneficial than all the others, $s_\beta \gg |s_{\alpha \ne \beta}|$,
and under the additional assumptions of $N s_\beta \gg 1$, $N s_\beta x^{\beta}_{0} \gg 1$, and $N |s_{\alpha \ne \beta}| \ll 1$,
Eq.~\eqref{Eq:fluxes:final:M2} becomes:
\begin{equation} \label{Eq:fluxes:final:M6}
P_{\text{fix}}^{\beta} \simeq 1,
\end{equation}
such that the strongly beneficial allele $A_\beta$ is nearly guaranteed to fix in the population. 

\subsection{Perturbative expansion of fixation probabilities}
The fixation probabilities satisfy the following equation in the absence of mutations, a generalization of Eq.~\eqref{FP:fixprob} to an arbitrary number of alleles:
\begin{equation} \label{FP:fixprob:M}
\left[ s^\alpha (\mathbf{x}) \partial_\alpha + \frac{1}{2N} g^{\alpha \beta} (\mathbf{x}) \partial_{\alpha} \partial_{\beta} \right] P_{\text{fix}}^{\gamma} (\mathbf{x}; \mathbf{s}) \equiv
\Lop_s P_{\text{fix}}^{\gamma} (\mathbf{x}; \mathbf{s}) = 0,
\end{equation}
where $\Lop_s$ is the backward operator, $\mathbf{s} = (s_1, \ldots, s_{M-1})$, 
and $s^\alpha = x^\alpha (s_\alpha - \langle s \rangle)$, with $\langle s \rangle = \sum_{\beta} x^\beta s_\beta$;
we have dropped the $0$ subscript from the initial allele frequencies for brevity.
Note that $P_{\text{fix}}^{M}$ is determined by Eq.~\eqref{Eq:Pfix:norm}.

We now develop an expansion of the fixation probabilities in powers of $N s_\alpha$:
\begin{equation} \label{Pfix:expan}
P_{\text{fix}}^{\gamma} = P_{\text{fix,0}}^{\gamma} + P_{\text{fix,1}}^{\gamma} + P_{\text{fix,2}}^{\gamma} + \mathcal{O} ((N s_\alpha)^3),
\end{equation}
using Eq.~\eqref{Eq:fluxes:final:M2} to construct plausible initial guesses for each expansion term.

For simplicity, we focus first on the $M=3$ case, for which Eq.~\eqref{FP:fixprob:M} becomes:
\begin{eqnarray} \label{FP:fixprob:M3}
[ g^{11} \partial^2_1 &+& 2 g^{12} \partial_1 \partial_2 + g^{22} \partial^2_2 ] P_{\text{fix}}^{\gamma} = \nonumber \\
&-&2N [ x^1 (s_1 - \langle s \rangle) \partial_1 + x^2 (s_2 - \langle s \rangle) \partial_2 ] P_{\text{fix}}^{\gamma},
\end{eqnarray}
with $\langle s \rangle = x^1 s_1 + x^2 s_2$. The first-order expansion of Eq.~\eqref{Eq:fluxes:final:M2} suggests $P_{\text{fix,0}}^{\gamma} = x^\gamma$ (cf. Eq.~\eqref{Eq:fluxes:final:M3}),
in accordance with the neutral result. Note that $P_{\text{fix,0}}^{\gamma}$ satisfies Eq.~\eqref{FP:fixprob:M3} to the zeroth order in $Ns_1,Ns_2$, with the appropriate boundary conditions at the vertices
of the simplex.

The second-order expansion of Eq.~\eqref{Eq:fluxes:final:M2} results in
\begin{eqnarray} \label{Constr:M3:2}
s_1 P_{\text{fix,1}}^{1} &+& s_2 P_{\text{fix,1}}^{2} \\
&=& N s_1 x^1 (s_1 - \langle s \rangle) + N s_2 x^2 (s_2 - \langle s \rangle), \nonumber
\end{eqnarray}
which strongly suggests
\begin{equation} \label{Fixprob:M3:1}
P_{\text{fix,1}}^{\gamma} = N x^\gamma (s_\gamma - \langle s \rangle).
\end{equation}
Indeed, $P_{\text{fix,1}}^{\gamma}$ defined by Eq.~\eqref{Fixprob:M3:1} satisfies Eq.~\eqref{FP:fixprob:M3} to the first order in $Ns_1,Ns_2$ and is consistent with
a recent independent derivation, which relied on a perturbative expansion of the equivalent of Eq.~\eqref{FP:fixprob:M} for the Moran process~\citep{Braga2026}.
Note that $P_{\text{fix,1}}^{\gamma} = 0$ at the simplex vertices. 

Finally, the third-order expansion of Eq.~\eqref{Eq:fluxes:final:M2} results in
\begin{eqnarray} \label{Constr:M3:3}
s_1 P_{\text{fix,2}}^{1} &+& s_2 P_{\text{fix,2}}^{2} \\
&=& N ^2 s_1^2 x^1 (s_1 - \langle s \rangle) - \frac{2}{3} N^2 s_1 x^1 (s_1^2 - \langle s \rangle^2) \nonumber \\
&+& [ 1 \leftrightarrow 2], \nonumber
\end{eqnarray}
yielding
\begin{eqnarray} \label{Fixprob:M3:2}
P_{\text{fix,2}}^{\gamma} &=& N ^2 s_\gamma x^\gamma (s_\gamma - \langle s \rangle) \\
&-& \frac{2}{3} N^2 x^\gamma (s_\gamma^2 - \langle s \rangle^2) + \tilde{P}_{\text{fix,2}}^{\gamma}, \nonumber
\end{eqnarray}
where the ``residual'' contributions $\tilde{P}_{\text{fix,2}}^{\gamma}$ satisfy $s_1 \tilde{P}_{\text{fix,2}}^{1} + s_2 \tilde{P}_{\text{fix,2}}^{2} = 0$
and therefore must contain antisymmetric combinations of $s_1$ and $s_2$. Note that $\tilde{P}_{\text{fix,1}}^{\gamma} = 0$ in Eq.~\eqref{Fixprob:M3:1} because
no antisymmetric combinations satisfying $s_1 \tilde{P}_{\text{fix,1}}^{1} + s_2 \tilde{P}_{\text{fix,1}}^{2} = 0$ could be constructed at that order.

Here, symmetry considerations, including the fact that $\tilde{P}_{\text{fix,2}}^{2}$ can be obtained from $\tilde{P}_{\text{fix,2}}^{1}$ by $1 \leftrightarrow 2$, lead to the following ansatz:
\begin{equation} \label{Fixprob:M3:2res}
\tilde{P}_{\text{fix,2}}^{1} = C N^2 (s_1 - s_2) s_2 x^1 x^2,
\end{equation}
where $C$ is a constant which is found to be equal to $1/3$ by plugging Eq.~\eqref{Fixprob:M3:2} with $\gamma = 1$ into Eq.~\eqref{FP:fixprob:M3}
and retaining the second-order terms in $s_1$ and $s_2$. 

The above formulation can be extended to an arbitrary number of alleles $M$, resulting in the following quadratic expansion of the fixation probabilities:
\begin{eqnarray} \label{Fixprob:fullM}
P_{\text{fix}}^{\gamma} &\simeq& x^\gamma + N x^\gamma (s_\gamma - \langle s \rangle) \\
&+& N ^2 s_\gamma x^\gamma (s_\gamma - \langle s \rangle) - \frac{2}{3} N^2 x^\gamma (s_\gamma^2 - \langle s \rangle^2) \nonumber \\
&+& \frac{1}{3} N^2 x^\gamma (s_\gamma \langle s \rangle - \langle s^2 \rangle), \nonumber
\end{eqnarray}
where $\langle s^2 \rangle = \sum_{\beta} x^\beta s_\beta^2$. These probabilities satisfy Eq.~\eqref{FP:fixprob:M} if the third-order terms with respect to the selection coefficients are neglected.

\subsection{Exact recursive series expansion of fixation probabilities}
Inspired by Eq.~\eqref{Fixprob:fullM}, we present the fixation probability of allele $\gamma$ as
\begin{equation}
P^\gamma_{\fix}
=
x^\gamma \sum_{k=0}^{\infty} N^k Q_k^\gamma,
\label{eq:main-expansion}
\end{equation}
where coefficients $Q_k^\gamma$ are homogeneous polynomials of degree $k$ in the selection coefficients, computed recursively as explained below ($Q_0^\gamma = 1,~\forall \gamma$).
Eq.~\eqref{eq:main-expansion} constitutes the second major theoretical result of this work.



We let
$a \equiv a_\gamma=s_\gamma - \avg{s}$
and define the centered moments:
\begin{equation}
\mu_n = \avg{(s-\avg{s})^n}
= \sum_{i=1}^M x^i (s_i-\avg{s})^n,
\label{eq:mu-def}
\end{equation}
where $n = 0, 1, 2, \dots$; note that $\mu_0=1$, $\mu_1=0$.
We also introduce
\begin{equation}
p_n = \avg{(s_\gamma-s)^n} = \sum_{j=0}^n (-1)^j \binom{n}{j} a^{n-j} \mu_j.
\label{eq:p-def}
\end{equation}
For example, $p_1 = a$, $p_2 = a^2+\mu_2$, $p_3 = a^3+3a\mu_2-\mu_3$, and so on.


Next, we define a partition $\lambda \vdash k$ as a way of writing $k$ as a sum of positive integers, disregarding their order. For example,
\begin{equation*}
4=4=3+1=2+2=2+1+1=1+1+1+1.
\end{equation*}
If $\lambda=(\lambda_1,\ldots,\lambda_\ell)$ is a partition, its length is given by $\ell(\lambda) = \ell$, the number of positive integers in the partition. 
For each partition $\lambda$, we define $p_\lambda=p_{\lambda_1} \cdots p_{\lambda_\ell}$.
For instance, $p_{3,1}=p_3 p_1$, $p_{2,2,1}=p_2^2 p_1$.


The recursion involves two operators, denoted $\Dop$ and $\Sop$. In the $p_\lambda$ basis, their actions are as follows:
\begin{equation}
\Dop p_\lambda
=
-\ell(\lambda)\bigl[\ell(\lambda)+1\bigr]p_\lambda
+ 2 \Mop p_\lambda,
\label{eq:D-on-p}
\end{equation}
where $\Mop$ is the pair-merging operator -- it chooses two factors in $p_\lambda$ and merges their indices:
\begin{equation}
\Mop p_{\lambda_1,\ldots,\lambda_\ell}
=
\sum_{1\leq i<j\leq \ell}
 p_{\lambda_i+\lambda_j}
 \prod_{m\neq i,j}p_{\lambda_m}.
\label{eq:M-def}
\end{equation}
Examples are
\begin{align}
\Mop p_r &= 0,
\label{eq:M1} \\
\Mop p_{r,s} &= p_{r+s}, \nonumber \\
\Mop p_{r,s,t} &= p_{r+s}p_t+p_{r+t}p_s+p_{s+t}p_r. \nonumber
\end{align}
Repeated factors are counted with multiplicity.  For example, $\Mop(p_1^4)=6p_2p_1^2$.

The selection operator is defined as:
\begin{equation}
\Sop p_{\lambda_1,\ldots,\lambda_\ell}
=
(\ell+1)p_{\lambda\cup 1}
-
\sum_{i=1}^{\ell}
 p_{\lambda_i+1}\prod_{m\neq i}p_{\lambda_m},
\label{eq:S-on-p}
\end{equation}
where $\lambda\cup 1$ signifies appending $1$ to the partition. For instance,
\begin{align} \label{eq:Spn}
\Sop p_{n} &= 2p_np_1-p_{n+1}, \\
\Sop p_{1,1} &= 3p_1^3-2p_2p_1. \nonumber
\end{align}


The $p_\lambda$ basis is not diagonal under $\Dop$ because of the merging term in Eq.~\eqref{eq:D-on-p}.  We therefore introduce a basis $Y_\lambda$ satisfying
\begin{equation}
\Dop Y_\lambda
=
-\ell(\lambda)\bigl[\ell(\lambda)+1\bigr]Y_\lambda.
\label{eq:Y-eigen}
\end{equation}
We choose the monic normalization: 
\begin{equation}
Y_\lambda=p_\lambda+\text{terms of lower length}.
\label{eq:Y-monic}
\end{equation}
To construct $Y_\lambda$, we write 
\begin{equation}
Y_\lambda=F_\ell+F_{\ell-1}+\cdots+F_1,
\label{eq:Y-F-decomp}
\end{equation}
where $F_j$ is a linear combination of $p$-monomials of length $j$.  We start from $F_\ell=p_\lambda$ and then use
\begin{equation}
F_j=
-\frac{2}{(\ell-j)(\ell+j+1)}\,
\Mop F_{j+1},
\qquad
j=\ell-1,\ldots,1.
\label{eq:F-recursion}
\end{equation}
Equivalently,
\begin{equation}
Y_\lambda
=
\sum_{q=0}^{\ell-1}
\frac{(-2)^q(2\ell-q)!}{q!(2\ell)!}
\Mop^q p_\lambda.
\label{eq:Y-closed}
\end{equation}

The first few basis elements are $Y_{n}=p_n$, $Y_{r,s}=p_rp_s-\frac12p_{r+s}$, and so on.
For example, $Y_{1,1,1} = p_1^3-p_1p_2+\frac15p_3$.


Finally, the perturbative recursion for $Q_{k}$ (we omit $\gamma$ for brevity) is given by:
\begin{equation}
\Dop Q_{k+1}=-2 \Sop Q_k,
\label{eq:Q-recursion}
\end{equation}
where $Q_k$ can be expanded as:
\begin{equation}
Q_k=\sum_{\lambda\vdash k}q^{(k)}_\lambda Y_\lambda.
\label{eq:Q-Y-expansion}
\end{equation}
Because $Y_\lambda$ diagonalizes $\Dop$, Eq.~\eqref{eq:Q-recursion} becomes the scalar rule:
\begin{equation}
q^{(k+1)}_\eta
=
\frac{2}{\ell(\eta)\bigl[\ell(\eta)+1\bigr]}
\,[Y_\eta] \Sop Q_k,
\label{eq:q-recursion}
\end{equation}
where $[Y_\eta]G$ means the coefficient of $Y_\eta$ after re-expanding $G$ in the $Y$-basis.

The above considerations can be summarized as the following recursive algorithm for computing $Q_k$:
\begin{enumerate}
\item Start from $Q_0=1$.
\item Suppose that $Q_k$ is known in the $Y$-basis and that $Y_\lambda$ functions are expressed in the $p$-basis.
\item Apply $\Sop$ using Eq.~\eqref{eq:S-on-p}. This gives $\Sop Q_k$ in the $p_\lambda$ basis.
\item Re-expand $\Sop Q_k$ in the $Y_\lambda$ basis by triangular back-substitution.
\item For each $Y_\eta$, multiply its coefficient by $2/[\ell(\eta)(\ell(\eta)+1)]$ to obtain $Q_{k+1}$.
\end{enumerate}


The triangular back-substitution proceeds as follows: suppose that after applying $\Sop$ one has
\begin{equation}
G=\sum_\lambda g_\lambda p_\lambda.
\label{eq:G-p}
\end{equation}
We want to re-express $G$ in the $Y$-basis:
\begin{equation}
G=\sum_\lambda c_\lambda Y_\lambda.
\label{eq:G-Y}
\end{equation}
Due to Eq.~\eqref{eq:Y-monic}, this 
change of basis is triangular -- the coefficient of the largest-length $p_\lambda$ term in \eqref{eq:G-p} is the coefficient of the corresponding $Y_\lambda$.
Next, we subtract that multiple of $Y_\lambda$ and proceed to the next-largest length.

Explicitly, we set $G_{\rm rem}=G$ and repeat the following for $L=L_{\max},L_{\max}-1,\ldots,1$:
\begin{align}
&c_\lambda=[p_\lambda]G_{\rm rem},
\qquad \ell(\lambda)=L,
\label{eq:tri-c}\\
&G_{\rm rem} \leftarrow
G_{\rm rem} - \sum_{\ell(\lambda)=L}c_\lambda Y_\lambda. \nonumber
\end{align}
At the end, the coefficients $c_\lambda$ give the desired $Y$-basis expansion.


Finally, we provide a step-by-step illustration of the recursive algorithm.
Starting from $Q_0=1$, we obtain $\Sop Q_0=p_1=Y_{1}$, hence
\begin{equation}
Q_1=Y_{1}=a.
\label{eq:Q1-derive}
\end{equation}
Next, we consider
$\Sop Y_{1}=2Y_{1,1}$.
Since $\ell(1,1)=2$, Eq.~\eqref{eq:q-recursion} gives
$Q_2=\frac{2}{2\cdot 3}\,2Y_{1,1}=\frac23Y_{1,1}$.
Using $Y_{1,1}=\frac12(a^2-\mu_2)$, we obtain:
\begin{equation}
Q_2=\frac13(a^2-\mu_2).
\label{eq:Q2-amu}
\end{equation}
For the next step, we need
$Y_{1,1}=p_1^2-\frac12p_2$.
Using Eq.~\eqref{eq:S-on-p}, we get $\Sop p_{1,1} = 3p_1^3-2p_2p_1$ and $\Sop p_2 = 2p_2p_1-p_3$.
Therefore,
$\Sop Y_{1,1}=3p_1^3-3p_2p_1+\frac12p_3$.
Using the expansion of $Y_{1,1,1}$
in the $p$-basis,
we obtain
$\Sop Y_{1,1}=3Y_{1,1,1}-\frac1{10}Y_{3}$.
Thus,
$\Sop Q_2=2Y_{1,1,1}-\frac1{15}Y_{3}$.
Finally, applying Eq.~\eqref{eq:q-recursion} gives
\begin{equation}
Q_3=\frac13Y_{1,1,1}-\frac1{15}Y_{3} = -\frac13a\mu_2.
\label{eq:Q3-Y-derive}
\end{equation}

The recursion can be continued step-by-step; explicit expressions for $Q_0, \dots, Q_{10}$ are listed in \nameref{app:explicitQ}.
Note that with the $Q_1$ and $Q_2$ coefficients from Eqs.~\eqref{eq:Q1-derive} and \eqref{eq:Q2-amu}, the first three terms in Eq.~\eqref{eq:main-expansion} match Eq.~\eqref{Fixprob:fullM} exactly.
An efficient symbolic implementation of the above recursion is available at \url{https://github.com/afeigel/fixation-probabilities}.


Next, we consider some of the structural properties of the $Q$ coefficients.
We observe that all odd coefficients have the structure:
\begin{equation} \label{eq:Q2m1}
Q_{2m+1}=a\,R_{2m}(a,\mu_2,\ldots,\mu_{2m}),
\end{equation}
where \(R_{2m}\) is a homogeneous polynomial of total degree \(2m\) in the selection coefficients.
Thus, no pure-moment monomial appears in \(Q_{2m+1}\). For example, there
is no \(\mu_3\) term in \(Q_3\) and no \(\mu_5\), \(\mu_2\mu_3\) terms in
\(Q_5\). 

To prove this statement, we consider a selection reversal operator: $\mathbf{s} \mapsto -\mathbf{s}$. Eq.~\eqref{FP:fixprob:M} implies that
\begin{equation}
\Lop_s
\left[
e^{2Na_\gamma}P^\gamma_{\rm fix}(\mathbf{x}; -\mathbf{s})
\right]
=0.
\end{equation}
Since the function in brackets has the same fixation boundary values as
$P^\gamma_{\rm fix} (\mathbf{x}; \mathbf{s})$, we immediately obtain by uniqueness of the bounded solution of
the backward problem:
\begin{equation}
P^\gamma_{\rm fix}(\mathbf{x}; \mathbf{s}) = e^{2Na_\gamma} P^\gamma_{\rm fix}(\mathbf{x}; -\mathbf{s}).
\end{equation}

Next, we investigate the parity of the $Q$ coefficients under the selection reversal operator. 
Using Eq.~\eqref{eq:Q-Y-expansion} and observing that every $Y_\lambda$ in the $\lambda \vdash k$ sum is homogeneous of degree \(k\), we conclude that
$Q_k$ 
is also homogeneous of degree \(k\) in the selection coefficients, as claimed below Eq.~\eqref{eq:main-expansion}:
$Q_k(-\mathbf{s}) = (-1)^k Q_k(\mathbf{s})$.
Thus, \(Q_{2m}\) is even and \(Q_{2m+1}\) is odd under the selection reversal operator.

Defining $H_\gamma (N) = \sum_{k=0}^{\infty} N^k Q_k^\gamma$, we find that $H_\gamma (N) = e^{2Na_\gamma} H_\gamma (-N)$ and thus
$e^{-Na_\gamma} H_\gamma (N)$ is an even function of $N$. The vanishing of the odd powers of $N$ in this function then yields:
\begin{equation}
Q_{2m+1} = - \sum_{j=1}^{2m+1} \frac{(-a)^j}{j!} Q_{2m+1-j}.
\end{equation}
Every term on the right-hand side contains at least one explicit factor of $a$, thus proving Eq.~\eqref{eq:Q2m1}.






Finally, we observe empirically that
\begin{equation}
Q_{2m}
=
\frac{(-1)^{m+1}}{D_m}\,\mathcal P_{2m}
\end{equation}
and
\begin{equation}
Q_{2m+1}
=
\frac{(-1)^m}{D_m}\,a\,\mathcal R_{2m},
\end{equation}
where \(\mathcal P_{2m}\) and \(\mathcal R_{2m}\) are polynomials with integer
coefficients and  \(D_m\) is the common denominator.

The empirical formula for the denominator is given by
\begin{equation}
D_m=
\prod_{\substack{q\ \mathrm{odd\ prime}\\ q\leq 2m+1}}
q^{\left\lfloor 2m/(q-1)\right\rfloor}.
\end{equation}
For example, $D_1=3$, $D_2=3^2\cdot5=45$, $D_3=3^3\cdot5\cdot7=945$, and $D_4=3^4\cdot5^2\cdot7=14175$.
The denominators grow rapidly with $m$ and thus play a key role in the convergence of the series in Eq.~\eqref{eq:main-expansion}.


\subsection{Numerical tests of the recursive fixation probability formula}

In principle, Eq.~\eqref{eq:main-expansion} can be used to compute fixation probabilities for a set of mutations characterized by arbitrary selection coefficients. However, in practice
the series expansion is limited by the rapidly increasing length and complexity of the higher-order terms. 
We find that a reasonable balance between numerical accuracy and computational efficiency is achieved by retaining terms with $k = (0, \dots, k_\text{max}=36)$
in Eq.~\eqref{eq:main-expansion}, which allows the series to converge if $N |s_\text{max} - s_\text{min}| \le 2.5$, where $s_\text{min}$ and $s_\text{max}$ are the smallest and the largest selection coefficients,
respectively. This estimate is empirical; theoretical estimates of the radius of convergence are the subject of future work.

\begin{figure}[!htbp]
\centering
\includegraphics[width=1.05\linewidth]{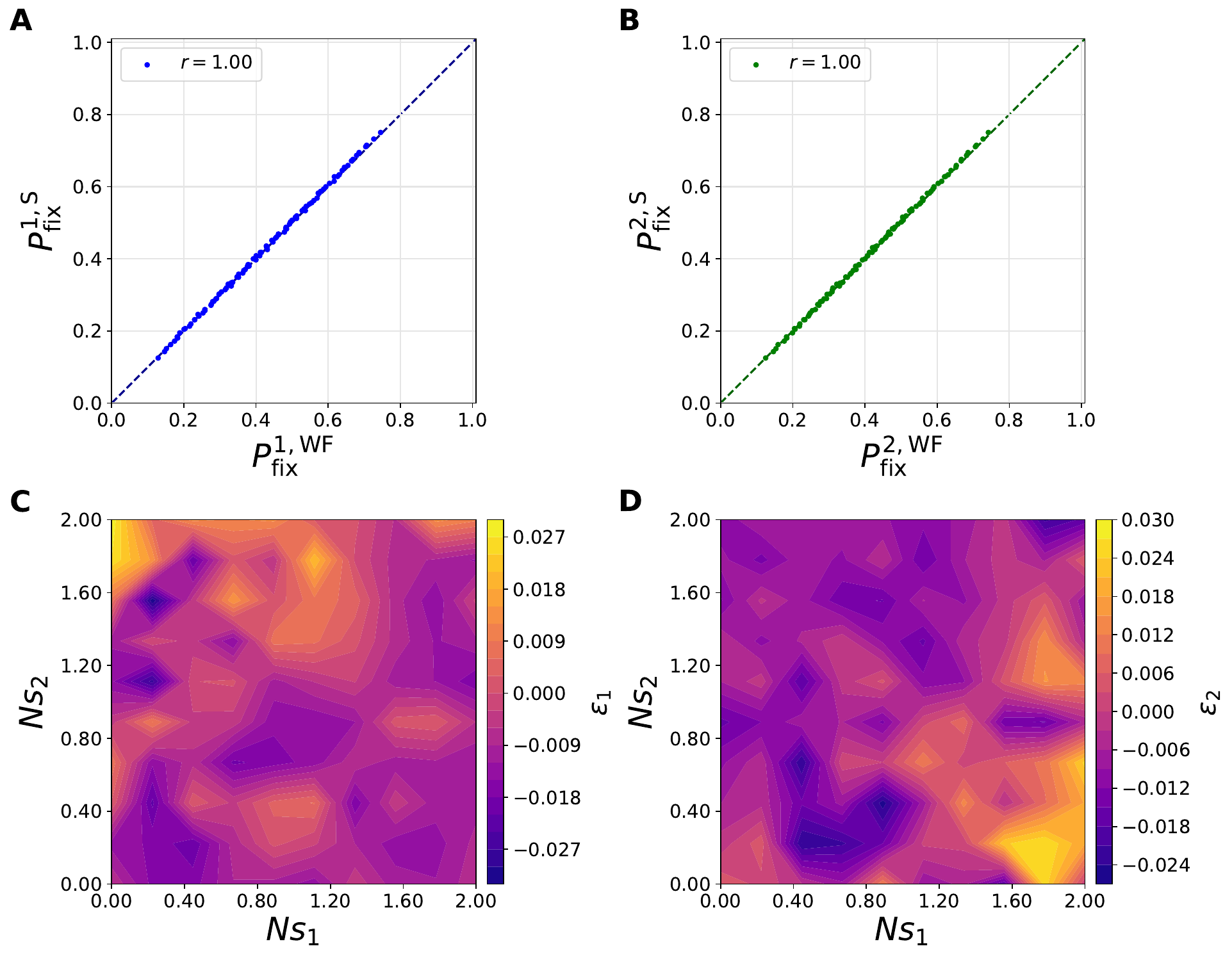}
\caption{ \textbf{Comparison of the series expansion and Wright-Fisher fixation probabilities in a three-allele system ($\mathbf{M=3}$).}
Both $s_1$ and $s_2$ selection coefficients were scanned over $10$ equidistant values in the $[0.00,0.02]$ range, resulting in $10^2$ unique combinations.
Wright-Fisher fixation probabilities were computed using $2 \times 10^4$ independent runs, each run resulting in the fixation of one of the alleles.
The initial allele frequencies are given by $x^i_0 = 1/3,~\forall i$; the population size is $N=100$.
A: a scatter plot of allele $1$ series expansion fixation probability, $P_\fix^{\mathrm{1,S}}$ (Eq.~\eqref{eq:main-expansion} with $k_\text{max} = 36$), vs.
allele $1$ Wright-Fisher fixation probability,  $P_\fix^{1,\mathrm{WF}}$. The dashed blue line is $y=x$; $r$ is the linear correlation coefficient.
B: Same as (A), but for allele $2$ fixation probabilities.
C: A contour plot of the relative errors between allele $1$ series expansion and Wright-Fisher fixation probabilities, $\varepsilon_1 = (P_\fix^{1,\mathrm{WF}} - P_\fix^{1,\mathrm{S}})/P_\fix^{1,\mathrm{WF}}$.
D: A contour plot of the relative errors between allele $2$ series expansion and Wright-Fisher fixation probabilities, $\varepsilon_2 = (P_\fix^{2,\mathrm{WF}} - P_\fix^{2,\mathrm{S}})/P_\fix^{2,\mathrm{WF}}$.
}
\label{fig:WFcomparo}
\end{figure}

We have tested the $k_\text{max} = 36$ series expansion against Wright-Fisher simulations (Fig.~\ref{fig:WFcomparo}) and the numerically integrated Langevin equation, Eq.~\eqref{Lng:eq:1} (Fig.~S1). First, we have focused
on the fixation probabilities of alleles $1$ and $2$ in the three-allele system ($M=3$) with the population size-rescaled selection coefficients $N s_1$ and $N s_2$ equidistantly spaced over the $[0.0,2.0]$ range.
In both Wright-Fisher and Langevin cases, we find nearly perfect correlation between theory and simulations, limited only by the numerical errors in the latter that are caused by factors such as the finite number of trajectories used to estimate fixation probabilities, and resulting in at most $3-4$\% relative errors overall (Figs.~\ref{fig:WFcomparo}C,D and S1C,D).

Next, we have tested our series expansion against Wright-Fisher and Langevin simulations in systems with $M=5,10,20$ alleles (Fig.~S2). Again, we find nearly perfect correlation between theory and both types
of simulations, lending further numerical support to the idea that Eq.~\eqref{eq:main-expansion} holds generically. Thus, we focus on the series expansion-based results in the rest of the manuscript.

\begin{figure}[!htbp]
\centering
\includegraphics[width=1.05\linewidth]{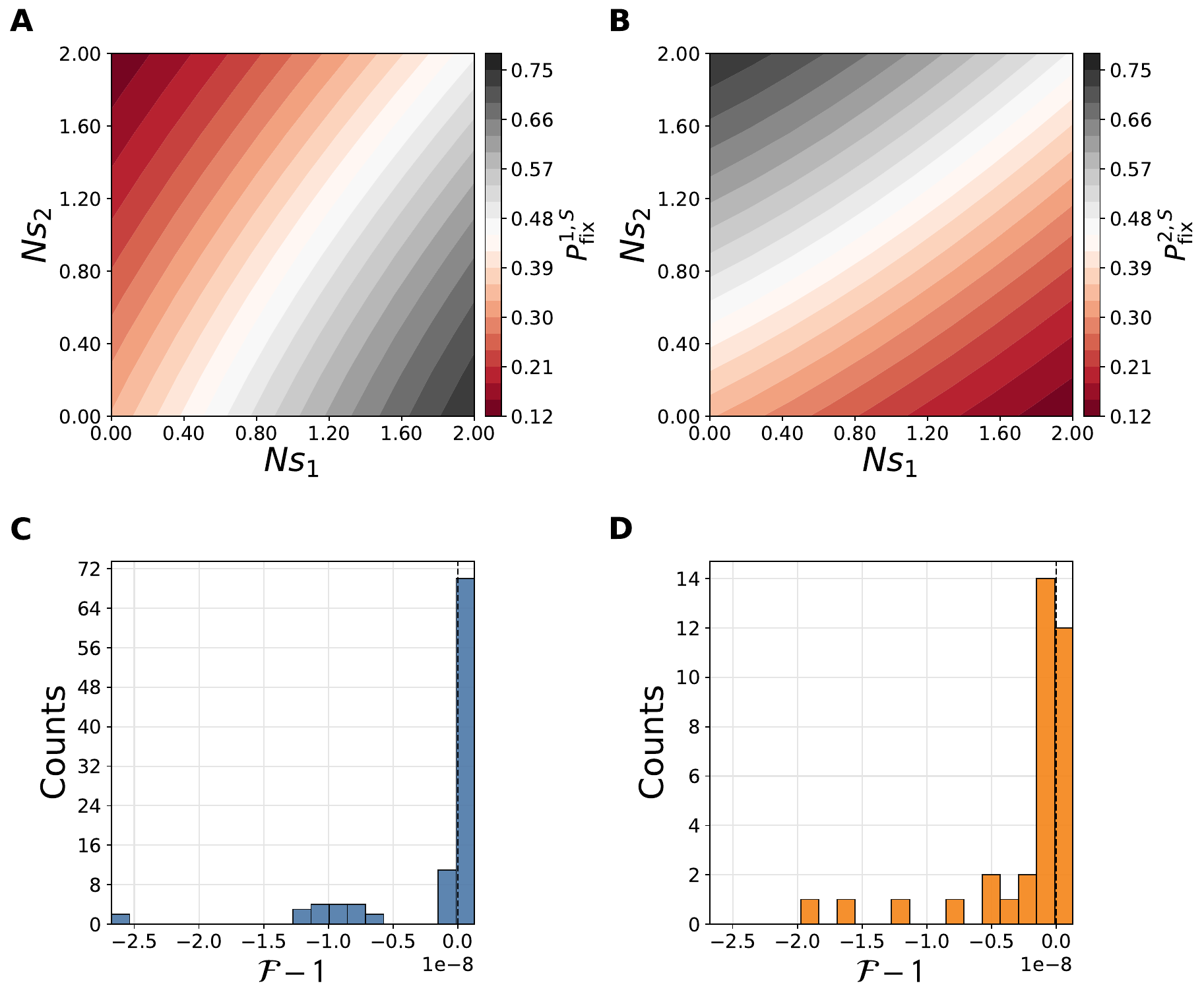}
\caption{ \textbf{Fixation probabilities and the fluctuation theorem constraint.}
A: A contour plot of the fixation probabilities $P_\fix^{1,\mathrm{S}}$ of allele $1$ in a three-allele system ($M=3$). The fixation probabilities were computed
using Eq.~\eqref{eq:main-expansion} with $k_\text{max} = 36$. As in Fig.~\ref{fig:WFcomparo}, $s_1$ and $s_2$ selection coefficients were scanned over $10$ equidistant values in the $[0.00,0.02]$ range,
resulting in $10^2$ unique combinations. The initial allele frequencies are given by $x^i_0 = 1/3,~\forall i$; the population size is $N=100$.
B: Same as (A), but for the fixation probabilities $P_\fix^{2,\mathrm{S}}$ of allele $2$.
C: A histogram of the fluctuation theorem constraint, $\mathcal{F} (P_{\text{fix}}^{1}, P_{\text{fix}}^{2}; s_1, s_2; x^{1}_{0}, x^{2}_{0}) - 1$ (cf. Eq.~\eqref{Eq:fluxes:final:M2}), for the $100$
sets of selection coefficients and fixation probabilities displayed in panels (A) and (B).
D: A histogram of the fluctuation theorem constraint, $\mathcal{F} (\{ P_{\text{fix}}^{\alpha} \}, \{ s_\alpha \}, \{ x^{\alpha}_{0} \}) - 1$ (cf. Eq.~\eqref{Eq:fluxes:final:M2}),
for the $35$ sets of multi-allele fixation probabilities shown in Fig.~S2.
}
\label{fig:Fluct}
\end{figure}

Fig.~\ref{fig:Fluct}A,B shows the heat maps of fixation probabilities $P_\fix^{1,\mathrm{S}}$ and $P_\fix^{2,\mathrm{S}}$ in the $M=3$ system (same data as in Fig.~\ref{fig:WFcomparo}).
These fixation probabilities satisfy the fluctuation theorem in Eq.~\eqref{Eq:fluxes:final:M2} within numerical precision (Fig.~\ref{fig:Fluct}C).
The fluctuation theorem is also satisfied by the series expansion fixation probabilities for $M=5,10,20$ that were used in Fig.~S2 (Fig.~\ref{fig:Fluct}D).
Taken together, these observations provide an independent verification of the fluctuation theorem constraint.

\subsection{Fixation probabilities in the presence of clonal interference}

\begin{figure}[!htbp]
\centering
\includegraphics[width=1.05\linewidth]{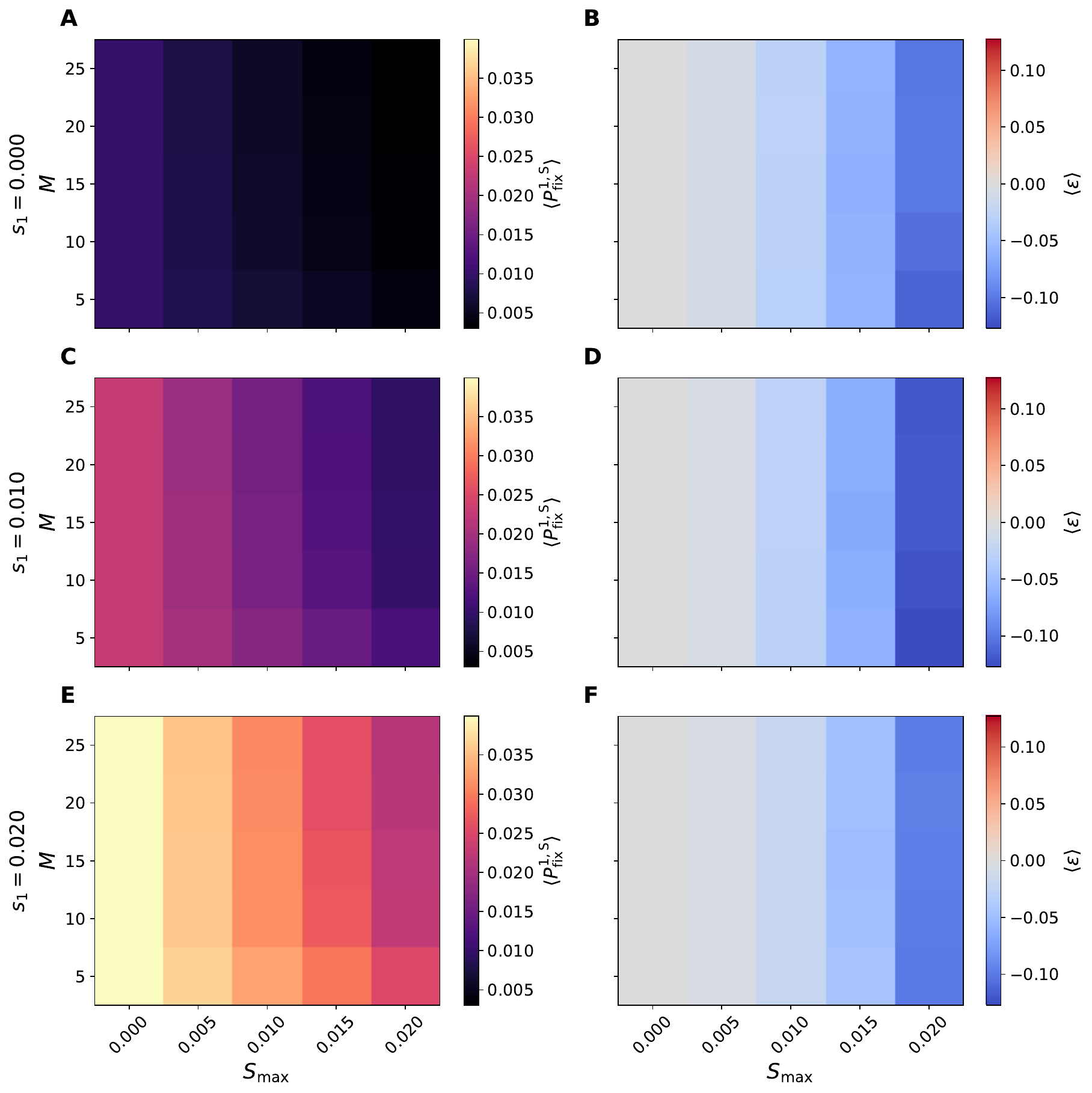}
\caption{ \textbf{Fixation probabilities of the target mutation in the presence of interfering beneficial mutations.}
Fixation probabilities of the target mutation with the selection coefficient $s_1$, $P_\fix^{1,\mathrm{S}}$, were computed using Eq.~\eqref{eq:main-expansion} with $k_\text{max} = 36$, for $5$ systems with
$M = \{5,10,15,20,25\}$.
The initial allele frequencies were set to $x^1_0 = 1/N$, $x^i_0 = (1 - 1/N)/(M -1),~i = 2, \dots, M$; the population size is $N=100$.
For each value of $M$, selection coefficients $s_2, \dots, s_{M-1}$ were uniformly sampled in the $[0.00, S_{\max})$ range, with $s_M = 0.0$
as required by our formalism. For each combination of $M$ and $S_{\max}$, the $P_\fix^{1,\mathrm{S}}$ values in panels (A), (C), (E) were averaged over $100$ independently
sampled sets of selection coefficients $s_2, \dots, s_{M-1}$. Panels (A), (C), (E) show these average fixation probabilities, $\langle P_\fix^{1,\mathrm{S}} \rangle$, for $s_1 = 0.00, 0.01, 0.02$, respectively.
Panels (B), (D), (F) show relative errors between the exact fixation probabilities and the mean-field Kimura approximation, Eq.~\eqref{Eq:Kimura:eff}:
$\varepsilon = (P_\fix^{1,\mathrm{S}} - P_\fix^{1,\mathrm{K}})/P_\fix^{1,\mathrm{S}}$, where $P_\fix^{1,\mathrm{K}}$ denotes Kimura's fixation probability.
For each combination of $M$ and $S_{\max}$, the relative errors were averaged over $100$ independent realizations of selection coefficients used in panels (A), (C), (E).
}
\label{fig:Tinter}
\end{figure}

We consider the fate of a novel mutant, arbitrarily labeled as allele $1$, which is characterized by a beneficial selection coefficient $s_1$ and initially appears in the population as a single copy.
In the absence of other mutations, the mutant competes with the rest of the population carrying the ancestral genotypes, and its fixation probability is given by Kimura's formula,
Eq.~\eqref{Eq:fluxes:final}. However, asexual evolution is typically characterized by clonal interference, in which several distinct lineages simultaneously spread and compete
in the population. To model this scenario, we study the fixation probability of a single-copy neutral or beneficial mutant which appears on the background of $M-1$ established 
lineages (Figs.~\ref{fig:Tinter} and S3; for simplicity, we assign equal frequencies to all established lineages, although this is not essential for our formalism). The selection
coefficients of the $M-2$ established mutant lineages are randomly chosen in the $[0.00, S_{\max})$ range, setting up a standard clonal interference scenario.

We observe that the dependence on the number of interfering alleles is rather weak. As expected, the fixation probability of the novel mutant decreases as the interfering mutations become more
beneficial and increases with the selection coefficient of the novel mutant itself (Figs.~\ref{fig:Tinter}A,C,E and S3A,C,E). However, these trends are also captured fairly accurately by a simple `mean-field' Kimura's
formula in which the $M-1$ background lineages are represented by their average selection coefficient:
\begin{equation} \label{Eq:Kimura:eff}
P_\fix^{1,\mathrm{K}} = \frac{1 - e^{-2N x^1_0 (s_1 - \bar{s}_\text{bkg}) }}{1 - e^{-2N (s_1 - \bar{s}_\text{bkg}) }},
\end{equation}
where $\bar{s}_\text{bkg} = (1 - x^1_0)^{-1} \sum_{j=2}^M s_j x^j_0$.

\begin{figure}[!htbp]
\centering
\includegraphics[width=1.05\linewidth]{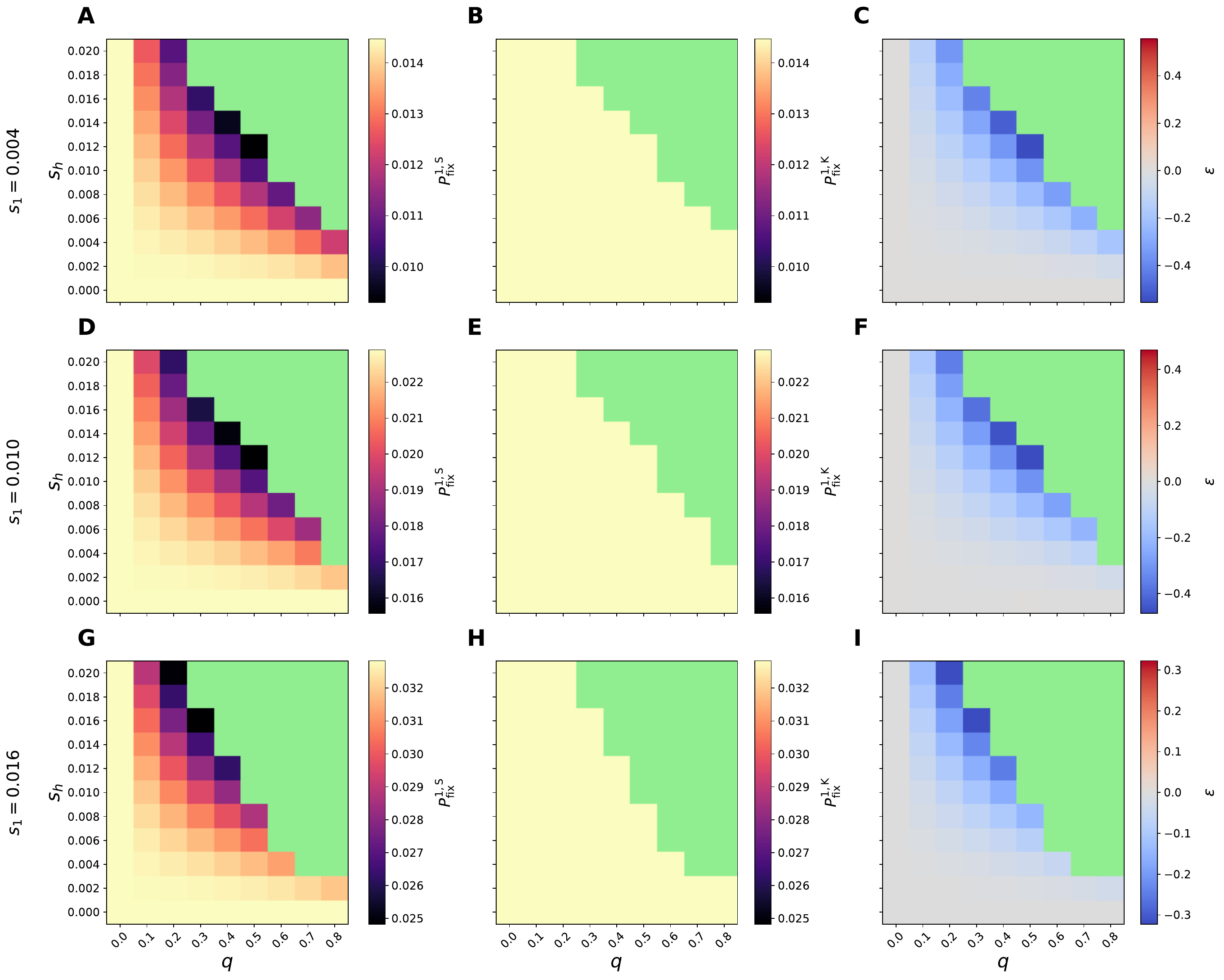}
\caption{ \textbf{Fixation probabilities of a beneficial target mutation in the presence of interfering beneficial and deleterious mutations with zero mean.}
Fixation probabilities $P_\fix^{1,\mathrm{S}}$ of the target mutation with the selection coefficient $s_1$ were computed using Eq.~\eqref{eq:main-expansion} with $k_\text{max} = 36$.
Kimura fixation probabilities $P_\fix^{1,\mathrm{K}}$ were computed using Eq.~\eqref{Eq:Kimura:eff} with $\bar{s}_\text{bkg} = 0$.
The initial allele frequencies were set to $(x^1_0, x^2_0, x^3_0) = (1/N, (1 - 1/N)q, (1 - 1/N)(1 - q))$.
The corresponding selection coefficients are $(s_1 - s_l, s_h - s_l, 0) = (s_1 + {q s_h}/{(1 - q)}, {s_h}/{(1 - q)}, 0)$; the population size is $N=100$.
Panels (A),(D),(G) show heatmaps of $P_\fix^{1,\mathrm{S}}$ as a function of $q$ and $s_h$, for $s_1 = (0.004, 0.010, 0.016)$, respectively.
Panels (B),(E),(H) show the corresponding heatmaps of $P_\fix^{1,\mathrm{K}}$.
Panels (C),(F),(I) show the relative errors between series expansion and Kimura fixation probabilities,
$\varepsilon = (P_\fix^{1,\mathrm{S}} - P_\fix^{1,\mathrm{K}})/P_\fix^{1,\mathrm{S}}$, for each value of $q$ and $s_h$.
Green areas indicate parameter combinations where at least one of the selection coefficients is greater than $0.025$ in magnitude,
preventing reliable convergence of the series expansion formula.
}
\label{fig:qscan}
\end{figure}

Indeed, as can be seen in Figs.~\ref{fig:Tinter}B,D,F and S3B,D,F, the absolute magnitude of the relative error between $P_\fix^{1,\mathrm{S}}$ and $P_\fix^{1,\mathrm{K}}$ is $\lesssim 10\%$, increasing with the
variance of the set of selection coefficients characterizing the pre-existing population. Moreover, the correction to the mean-field fixation probabilities is always negative. 
This is not surprising since the non-zero variance of the selection coefficients indicates the presence of fitter-than-average genotypes in the population; competition with those genotypes lowers the fixation probability of the target mutation. This effect cannot be captured by the mean-field treatment because it requires an \textit{a priori} average over the selection coefficients.

Qualitatively similar results are obtained when a single novel mutant competes with established background lineages that are either deleterious or neutral on average (Figs.~S4 and S5).
The expected trends in $\langle P_\fix^{1,\mathrm{S}} \rangle$ (weak dependence on $M$, increase as $s_1$ becomes more beneficial, decrease as the competing lineages become less deleterious)
are again captured by the mean-field Kimura's formula, with the absolute magnitude of the relative error $\lesssim 10\%$ as in the previous case. The error is always negative and tends to increase
when the fitness variance in the pre-existing population increases.

To focus on the effect of the fitness variance in the interfering mutations, we have followed the fate of a single individual with a target mutation introduced into a population with a mixture of
beneficial and deleterious mutations whose frequencies are balanced so that the mean is always zero: $q s_h + (1-q) s_l = 0$, where $s_h > 0, s_l < 0, q \in [0,1)$. In this case, the mean-field Kimura's formula in 
Eq.~\eqref{Eq:Kimura:eff} depends only on $s_1$, the selection coefficient of the target mutation, and is insensitive to the values of $q$ and $s_h$ (note that $s_l = - q s_h/(1 - q)$
due to the zero-mean condition).

We expect the discrepancy between exact and mean-field fixation probabilities to grow as the variance of the interfering mutations, $V(q,s_h)$, is increased.
Since $V(q,s_h) = q s_h^2/(1-q)$, we expect the relative error between exact and mean-field fixation probabilities to increase with both $q$ and $s_h$, staying approximately constant 
along the $s_h \sim \sqrt{\frac{1-q}{q}}$ contours. Indeed, the relative errors increase with $q$ and $s_h$, reaching $\sim 40\%$ for higher variance (Fig.~\ref{fig:qscan});
the contours of constant variance can also be discerned in Fig.~\ref{fig:qscan}A,D,G.

As before, the fixation probabilities are always overestimated by the mean-field Kimura's formula, leading to negative relative errors. If $s_h$ is fixed, $V(q,s_h)$ depends only on $q$, with the relative
errors reaching $\sim 40\%$ at the largest values of $q$ for which exact fixation probabilities could still be computed (Fig.~S6). Overall, we observe significant deviations between the exact
treatment and the simple mean-field solution if the effects of the mean and the variance of the interfering mutations are disentangled by construction.

\section{Discussion and conclusion}

In this work, we have described a novel approach to predicting fixation probabilities in evolving populations with $M$ simultaneously present alleles or genotypes.
First, we have used path integral treatment
of fluctuation theorems (non-perturbative equalities originally developed to describe mesoscopic systems in non-equilibrium statistical mechanics)
to derive an exact constraint obeyed by $M$ fixation probabilities.
In the $M=2$ case this constraint 
fully defines the fixation probability of a mutant competing for fixation with the wild-type, allowing us to recover the well-known Kimura's formula, Eq.~\eqref{Eq:fluxes:final}.

Second, we have developed a recursive formula for computing
fixation probabilities of the $M$ alleles, Eq.~\eqref{eq:main-expansion}, thus solving a central problem in mathematical population genetics that was first formulated in 1964~\citep{Crow2009}.
Although the length and the complexity of the expressions quickly grow with the number of terms retained in Eq.~\eqref{eq:main-expansion} (cf.~\nameref{app:explicitQ}), standard symbolic
manipulations have allowed us to automate the recursion, retaining the first $37$ terms in the series (there is a one-time cost of pre-computing and storing databases of intermediate expressions;
after that, fixation probability calculations only require on-the-fly construction and evaluation of the appropriate formulas).
This cutoff is sufficient for treating the differences in population size-rescaled selection coefficients of up to $2.5$, which includes the majority of systems of practical interest.

The recursive symbolic calculations can be used to replace numerical simulations of fixation probabilities based either on discrete evolutionary models of population dynamics such as Wright-Fisher or Moran~\citep{Ewens2004},
or on numerical integration of continuous-time stochastic differential equations (Figs.~\ref{fig:WFcomparo} and S1). Moreover, the resulting fixation probabilities obey the fluctuation theorem constraint with
numerical precision (Fig.~\ref{fig:Fluct}C,D), showing that our approach is self-consistent.

We have used the recursive fixation probabilities to track the fate of a novel target mutation in the presence of other deleterious or beneficial mutations -- a classic problem of clonal interference
in population genetics, thought to play a key role in microbial and viral evolution~\citep{Fisher1930,Muller1932,Gerrish1998}. We find that fairly accurate approximate fixation probabilities can be obtained
using a mean-field Kimura's formula, Eq.~\eqref{Eq:Kimura:eff}, which accounts for the average effect of the interfering mutations but is insensitive to their variance (Figs.~\ref{fig:Tinter} and S3,S4,S5).

However, when we remove the average effect
by ensuring that the mean selection coefficient of the interfering mutations is always kept at zero, we find $\lesssim 40\%$ discrepancies between our exact treatment and the mean-field Kimura's formula,
for biologically relevant rescaled selection strengths of $\lesssim 2.5$ (Figs.~\ref{fig:qscan} and S6). 
Thus, it is important to use the exact formalism if the variance of the interfering mutations is not negligibly small.
In all cases considered, exact results show the decrease in the target mutation fixation probabilities compared to the mean-field approach.

Our exact fixation probabilities can be used to construct microscopic evolutionary models in which competing genotypes are subject to mutation, selection, and genetic drift, similar to
previous treatments~\citep{Gerrish1998,Park2010,Schiffels2011} in which approximate fixation probability expressions were used to study, for example, adaptive evolution in linked genomes.
This is the primary direction of our future work.


\section*{Code and data availability}
The code and data required to compute fixation probabilities and reproduce all Figures in this paper are available at \url{https://github.com/afeigel/fixation-probabilities}.

\section*{Acknowledgments}
A.V.M. acknowledges financial and logistical support from the Center for Quantitative Biology, Rutgers University.
A.F. is grateful to the Interdisciplinary Computational Physics Laboratory at the Racah Institute of Physics for the allocation of computational resources and technical assistance.
ChatGPT (OpenAI) was used for routine algebraic manipulation of the fixation probability expansion, under the direction of the authors; all results were verified against independent numerical calculations.

\section*{Conflicts of interest}
Nothing to declare.

\section*{Author contributions}
E.Y. developed the theory and revised the manuscript.
A.F. revised the manuscript and assisted with numerical experiments.
A.V.M. developed the theory, designed the numerical experiments, drafted the manuscript, and supervised and administered the project.



\appendix



\section{Appendix A: Explicit expressions of the $Q_k$ coefficients for $k = 0, \dots, 10$}
\label{app:explicitQ}

\subsection{$Q$ coefficients in the $Y$-basis}

In this section, $1^m$ means $m$ repeated ones.  For example,
$Y_{3,1^4}=Y_{3,1,1,1,1}$.
\begin{align}
Q_0={}&1,
\label{eq:Q0-Y}\\ \nonumber
Q_1={}&Y_{1}, \\ \nonumber
Q_2={}&\frac23Y_{1^2}, \\ \nonumber
Q_3={}&\frac13Y_{1^3}-\frac1{15}Y_{3}, \\ \nonumber
Q_4={}&\frac{2}{15}Y_{1^4}-\frac5{63}Y_{3,1}+\frac1{63}Y_{2,2}, \\ \nonumber
Q_5={}&\frac{2}{45}Y_{1^5}-\frac7{135}Y_{3,1^2}+
\frac2{135}Y_{2,2,1}+\frac2{315}Y_{5}, \\ \nonumber
Q_6={}&\frac4{315}Y_{1^6}-
\frac4{165}Y_{3,1^3}+
\frac4{495}Y_{2,2,1^2}+
\frac1{135}Y_{5,1}
\notag\\ \nonumber
&-\frac1{315}Y_{4,2}+
\frac2{945}Y_{3,3}, \\ \nonumber
Q_7={}&\frac1{315}Y_{1^7}-
\frac{22}{2457}Y_{3,1^4}+
\frac8{2457}Y_{2,2,1^3}+
\frac7{1485}Y_{5,1^2}
\notag\\ \nonumber
&-\frac{13}{4455}Y_{4,2,1}+
\frac2{891}Y_{3,3,1}-
\frac1{4455}Y_{3,2,2}-
\frac1{1575}Y_{7},
\end{align}

\vspace*{-0.28cm}
\begin{equation*}
\hspace*{-0.5cm}
\begin{aligned}
Q_8={}&\frac2{2835}Y_{1^8}
-\frac{13}{4725}Y_{3,1^5}
+\frac1{945}Y_{2,2,1^4}
\\
&+\frac{44}{20475}Y_{5,1^3}
-\frac{1052}{675675}Y_{4,2,1^2}
+\frac{904}{675675}Y_{3,3,1^2}
\\
&-\frac8{32175}Y_{3,2,2,1}
+\frac4{225225}Y_{2,2,2,2}
-\frac{23}{31185}Y_{7,1}
\\
&+\frac2{6237}Y_{6,2}
-\frac4{10395}Y_{5,3}
+\frac1{6237}Y_{4,4},
\end{aligned}
\label{eq:Q8-Y}
\end{equation*}

\begin{equation*}
\begin{aligned}
Q_9={}&\frac2{14175}Y_{1^9}
-\frac1{1377}Y_{3,1^6}
+\frac2{6885}Y_{2,2,1^5}
\\
&+\frac{11}{14175}Y_{5,1^4}
-\frac{113}{184275}Y_{4,2,1^3}
+\frac{106}{184275}Y_{3,3,1^3}
\\
&-\frac1{6825}Y_{3,2,2,1^2}
+\frac1{61425}Y_{2,2,2,2,1}
-\frac{86}{184275}Y_{7,1^2}
\\
&+\frac{1786}{6081075}Y_{6,2,1}
-\frac{787}{2027025}Y_{5,3,1}
+\frac{16}{2027025}Y_{5,2,2}
\\
&+\frac{893}{6081075}Y_{4,4,1}
+\frac{239}{6081075}Y_{4,3,2}
-\frac{127}{6081075}Y_{3,3,3}
\\
&+\frac2{31185}Y_{9},
\end{aligned}
\label{eq:Q9-Y}
\end{equation*}

\begin{equation*}
\hspace*{0.78cm}
\begin{aligned}
Q_{10}={}&\frac4{155925}Y_{1^{10}}
-\frac{136}{807975}Y_{3,1^7}
+\frac8{115425}Y_{2,2,1^6}
\\
&+\frac{169}{722925}Y_{5,1^5}
-\frac{47}{240975}Y_{4,2,1^4}
+\frac{142}{722925}Y_{3,3,1^4}
\\
&-\frac{44}{722925}Y_{3,2,2,1^3}
+\frac2{240975}Y_{2,2,2,2,1^2}
-\frac1{4725}Y_{7,1^3}
\\
&+\frac{316}{2027025}Y_{6,2,1^2}
-\frac{41}{184275}Y_{5,3,1^2}
+\frac1{96525}Y_{5,2,2,1}
\\
&+\frac{158}{2027025}Y_{4,4,1^2}
+\frac1{25025}Y_{4,3,2,1}
-\frac1{225225}Y_{4,2,2,2}
\\
&-\frac1{45045}Y_{3,3,3,1}
+\frac2{675675}Y_{3,3,2,2}
+\frac{32}{429975}Y_{9,1}
\\
&-\frac{1382}{42567525}Y_{8,2}
+\frac{1621}{42567525}Y_{7,3}
-\frac{1382}{42567525}Y_{6,4}
\\
&+\frac{739}{42567525}Y_{5,5}.
\end{aligned}
\label{eq:Q10-Y}
\end{equation*}

\subsection{$Q$ coefficients in the $p$-basis}

\begin{align}
Q_0={}&1, \label{eq:Q0-p} \\ \nonumber
Q_1={}&p_1, \\ \nonumber
Q_2={}&\frac13\left(-p_2+2p_1^2\right), \\ \nonumber
Q_3={}&\frac13\left(-p_1p_2+p_1^3\right), \\ \nonumber
Q_4={}&\frac1{45}\left(-9p_1^2p_2-p_1p_3+p_4+2p_2^2+6p_1^4\right), \\ \nonumber
Q_5={}&\frac1{45}\left(-4p_1^3p_2-p_1^2p_3+p_1p_4+2p_1p_2^2+2p_1^5\right),
\end{align}

\vspace*{-0.5cm}
\begin{equation*}
\hspace*{0.88cm}
\begin{aligned}
Q_6={}&\frac1{945}\Bigl(
3p_1p_2p_3-5p_2p_4-30p_1^4p_2-12p_1^3p_3+12p_1^2p_4
\notag\\
&+24p_1^2p_2^2+2p_1p_5-2p_6+p_3^2-3p_2^3+12p_1^6
\Bigr), \\
Q_7={}&\frac1{945}\Bigl(
3p_1^2p_2p_3-5p_1p_2p_4-9p_1^5p_2-5p_1^4p_3+5p_1^3p_4
\notag\\
&+10p_1^3p_2^2+2p_1^2p_5-2p_1p_6+p_1p_3^2-3p_1p_2^3+3p_1^7
\Bigr),
\end{aligned}
\end{equation*}

\vspace*{-0.5cm}
\begin{equation*}
\hspace*{0.88cm}
\begin{aligned}
Q_8={}&\frac1{14175}\Bigl(
25p_1^3p_2p_3-42p_1^2p_2p_4-p_1p_3p_4-3p_1p_2^2p_3
\notag\\
&-4p_1p_2p_5-3p_3p_5+7p_2^2p_4+7p_2p_6-2p_2p_3^2
\notag\\
&-35p_1^6p_2-25p_1^5p_3+25p_1^4p_4+50p_1^4p_2^2+17p_1^3p_5
\notag\\
&-17p_1^2p_6+9p_1^2p_3^2-25p_1^2p_2^3-3p_1p_7+3p_8
\notag\\
&+2p_4^2+2p_2^4+10p_1^8
\Bigr), \\
Q_9={}&\frac1{14175}\Bigl(
10p_1^4p_2p_3-17p_1^3p_2p_4-p_1^2p_3p_4-3p_1^2p_2^2p_3
\notag\\
&-4p_1^2p_2p_5-3p_1p_3p_5+7p_1p_2^2p_4+7p_1p_2p_6-2p_1p_2p_3^2
\notag\\
&-8p_1^7p_2-7p_1^6p_3+7p_1^5p_4+14p_1^5p_2^2+7p_1^4p_5
\notag\\
&-7p_1^3p_6+4p_1^3p_3^2-10p_1^3p_2^3-3p_1^2p_7+3p_1p_8
\notag\\
&+2p_1p_4^2+2p_1p_2^4+2p_1^9
\Bigr),
\end{aligned}
\end{equation*}

\vspace*{-0.5cm}
\begin{equation*}
\hspace*{-0.43cm}
\begin{aligned}
Q_{10}={}&\frac1{467775}\Bigl(
5p_1p_2p_3p_4+13p_2p_3p_5+105p_1^5p_2p_3-181p_1^4p_2p_4
\notag\\
&-18p_1^3p_3p_4-54p_1^3p_2^2p_3-73p_1^3p_2p_5-58p_1^2p_3p_5
\notag\\
&+127p_1^2p_2^2p_4+129p_1^2p_2p_6-38p_1^2p_2p_3^2+2p_1p_4p_5
\notag\\
&+2p_1p_3p_6+4p_1p_2^3p_3+8p_1p_2^2p_5+13p_1p_2p_7
\notag\\
&-12p_4p_6+p_3^2p_4+10p_3p_7-12p_2^3p_4-21p_2^2p_6+4p_2^2p_3^2
\notag\\
&-23p_2p_8-10p_2p_4^2-54p_1^8p_2-56p_1^7p_3+56p_1^6p_4+112p_1^6p_2^2
\notag\\
&+76p_1^5p_5-76p_1^4p_6+47p_1^4p_3^2-105p_1^4p_2^3-56p_1^3p_7
\notag\\
&+56p_1^2p_8+37p_1^2p_4^2+36p_1^2p_2^4+10p_1p_9-p_1p_3^3
\notag\\
&+5p_5^2-2p_2^5-10p_{10}+12p_1^{10}
\Bigr).
\end{aligned}
\end{equation*}

\subsection{$Q$ coefficients in the original centered variables}

\begin{adjustwidth}{-2.38cm}{1cm}
\begin{align}
Q_0={}&1, \label{eq:Q0-amu} \\ \nonumber
Q_1={}&a, \\ \nonumber
Q_2={}&\frac13\left(a^2-\mu_2\right), \\ \nonumber
Q_3={}&-\frac13a\mu_2, \\ \nonumber
Q_4={}&\frac1{45}\left(-a^4-2a^2\mu_2-3a\mu_3+2\mu_2^2+\mu_4\right), \\ \nonumber
Q_5={}&\frac1{45}\left(3a^3\mu_2-3a^2\mu_3+a\mu_4+2a\mu_2^2\right),
\end{align}
\end{adjustwidth}

\vspace*{-0.4cm}
\begin{equation*}
\hspace*{-1.08cm}
\begin{aligned}
Q_6={}&\frac1{945}\Bigl(
2a^6+18a^4\mu_2-a^3\mu_3+3a^2\mu_2^2-13a^2\mu_4
\notag\\
&+11a\mu_2\mu_3+10a\mu_5-3\mu_2^3-5\mu_2\mu_4+\mu_3^2-2\mu_6
\Bigr), \\
Q_7={}&\frac1{945}\Bigl(
-10a^5\mu_2+20a^4\mu_3-11a^3\mu_2^2-20a^3\mu_4+11a^2\mu_2\mu_3
\notag\\
&+10a^2\mu_5-3a\mu_2^3-5a\mu_2\mu_4+a\mu_3^2-2a\mu_6
\Bigr),
\end{aligned}
\end{equation*}

\vspace*{-0.4cm}
\begin{equation*}
\hspace*{-0.88cm}
\begin{aligned}
Q_8={}&\frac1{14175}\Bigl(
-3a^8-60a^6\mu_2+58a^5\mu_3-40a^4\mu_2^2-2a^4\mu_4
\notag\\
&-3a^3\mu_2\mu_3-55a^3\mu_5-2a^2\mu_2^3+33a^2\mu_2\mu_4+5a^2\mu_3^2
\notag\\
&+53a^2\mu_6-13a\mu_2^2\mu_3-29a\mu_2\mu_5-21a\mu_7+2\mu_2^4+7\mu_2^2\mu_4
\notag\\
&-2\mu_2\mu_3^2+7\mu_2\mu_6-3\mu_3\mu_5+2\mu_4^2+3\mu_8
\Bigr), \\
Q_9={}&\frac1{14175}\Bigl(
21a^7\mu_2-63a^6\mu_3+29a^5\mu_2^2+105a^5\mu_4-58a^4\mu_2\mu_3
\notag\\
&-105a^4\mu_5+13a^3\mu_2^3+58a^3\mu_2\mu_4+63a^3\mu_6-13a^2\mu_2^2\mu_3
\notag\\
&-29a^2\mu_2\mu_5-21a^2\mu_7+2a\mu_2^4+7a\mu_2^2\mu_4-2a\mu_2\mu_3^2
\notag\\
&+7a\mu_2\mu_6-3a\mu_3\mu_5+2a\mu_4^2+3a\mu_8
\Bigr),
\end{aligned}
\end{equation*}


\vspace*{-0.4cm}
\begin{equation*}
\hspace*{-0.38cm}
\begin{aligned}
Q_{10}={}&\frac1{467775}\Bigl(
10a^{10}+350a^8\mu_2-619a^7\mu_3+343a^6\mu_2^2+557a^6\mu_4
\notag\\
&-336a^5\mu_2\mu_3-5a^5\mu_5+97a^4\mu_2^3+28a^4\mu_2\mu_4-41a^4\mu_3^2
\notag\\
&-545a^4\mu_6+15a^3\mu_2^2\mu_3+326a^3\mu_2\mu_5+40a^3\mu_3\mu_4+609a^3\mu_7
\notag\\
&+2a^2\mu_2^4-85a^2\mu_2^2\mu_4-19a^2\mu_2\mu_3^2-322a^2\mu_2\mu_6-3a^2\mu_3\mu_5
\notag\\
&-18a^2\mu_4^2-327a^2\mu_8+20a\mu_2^3\mu_3+79a\mu_2^2\mu_5+4a\mu_2\mu_3\mu_4
\notag\\
&+141a\mu_2\mu_7-3a\mu_3^3-24a\mu_3\mu_6+20a\mu_4\mu_5+90a\mu_9
\notag\\
&-2\mu_2^5-12\mu_2^3\mu_4+4\mu_2^2\mu_3^2-21\mu_2^2\mu_6+13\mu_2\mu_3\mu_5
\notag\\
&-10\mu_2\mu_4^2-23\mu_2\mu_8+\mu_3^2\mu_4+10\mu_3\mu_7-12\mu_4\mu_6+5\mu_5^2-10\mu_{10}
\Bigr).
\end{aligned}
\end{equation*}



\bibliography{refs_combined}



\renewcommand{\thefigure}{S\arabic{figure}}
\setcounter{figure}{0}

\renewcommand{\thetable}{S\arabic{table}}
\setcounter{table}{0}

\newpage
\section{Supplementary Figures}

\begin{figure}[!htbp]
\centering
\includegraphics[width=0.9\linewidth]{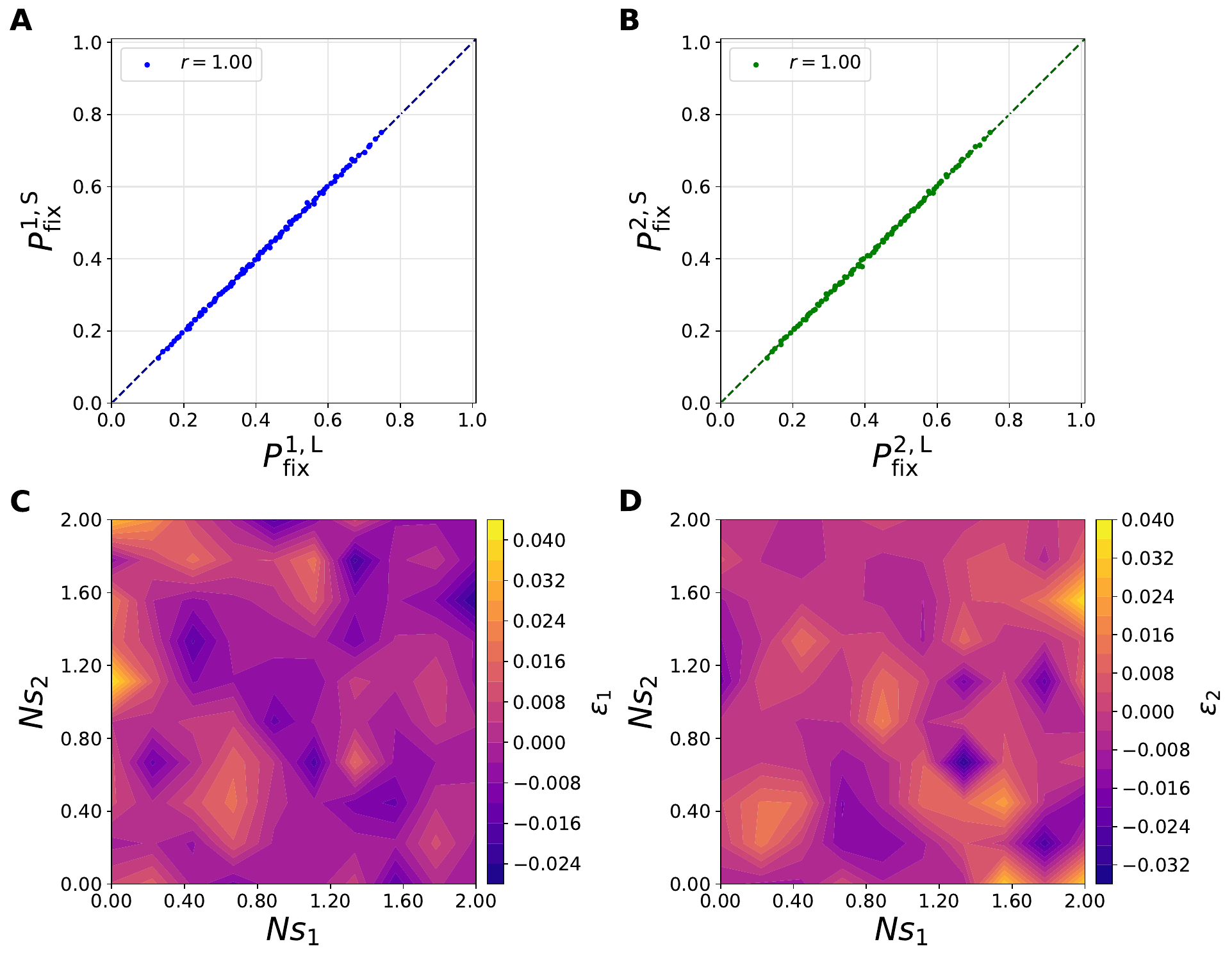}
\caption{ \textbf{Comparison of the series expansion and Langevin fixation probabilities in a three-allele system ($\mathbf{M=3}$).}
Both $s_1$ and $s_2$ selection coefficients were scanned over $10$ equidistant values in the $[0.00,0.02]$ range, resulting in $10^2$ unique combinations.
Langevin fixation probabilities were computed using $2 \times 10^4$ independent runs; in each run, Eq.~\eqref{Lng:eq:1} was solved numerically using the Euler–Maruyama method.
The simulations used the time step $dt = 1/20N$, where $N=100$ is the population size. The allele was considered extinct or fixed once its frequency reached $1/4N$ or $1 - 1/4N$, respectively.
The initial allele frequencies are given by $x^i_0 = 1/3,~\forall i$.
A: a scatter plot of allele $1$ series expansion fixation probability, $P_\fix^{1,\mathrm{S}}$ (Eq.~\eqref{eq:main-expansion} with $k_\text{max} = 36$), vs.
allele $1$ Langevin fixation probability,  $P_\fix^{1,\mathrm{L}}$. The dashed blue line is $y=x$; $r$ is the linear correlation coefficient.
B: Same as (A), but for allele $2$ fixation probabilities.
C: A contour plot of the relative errors between allele $1$ series expansion and Langevin fixation probabilities, $\varepsilon_1 = (P_\fix^{1,\mathrm{L}} - P_\fix^{1,\mathrm{S}})/P_\fix^{1,\mathrm{L}}$.
D: A contour plot of the relative errors between allele $2$ series expansion and Langevin fixation probabilities, $\varepsilon_2 = (P_\fix^{2,\mathrm{L}} - P_\fix^{2,\mathrm{S}})/P_\fix^{2,\mathrm{L}}$.
}
\label{fig:Lcomparo}
\end{figure}

\begin{figure}[!htbp]
\centering
\includegraphics[width=0.9\linewidth]{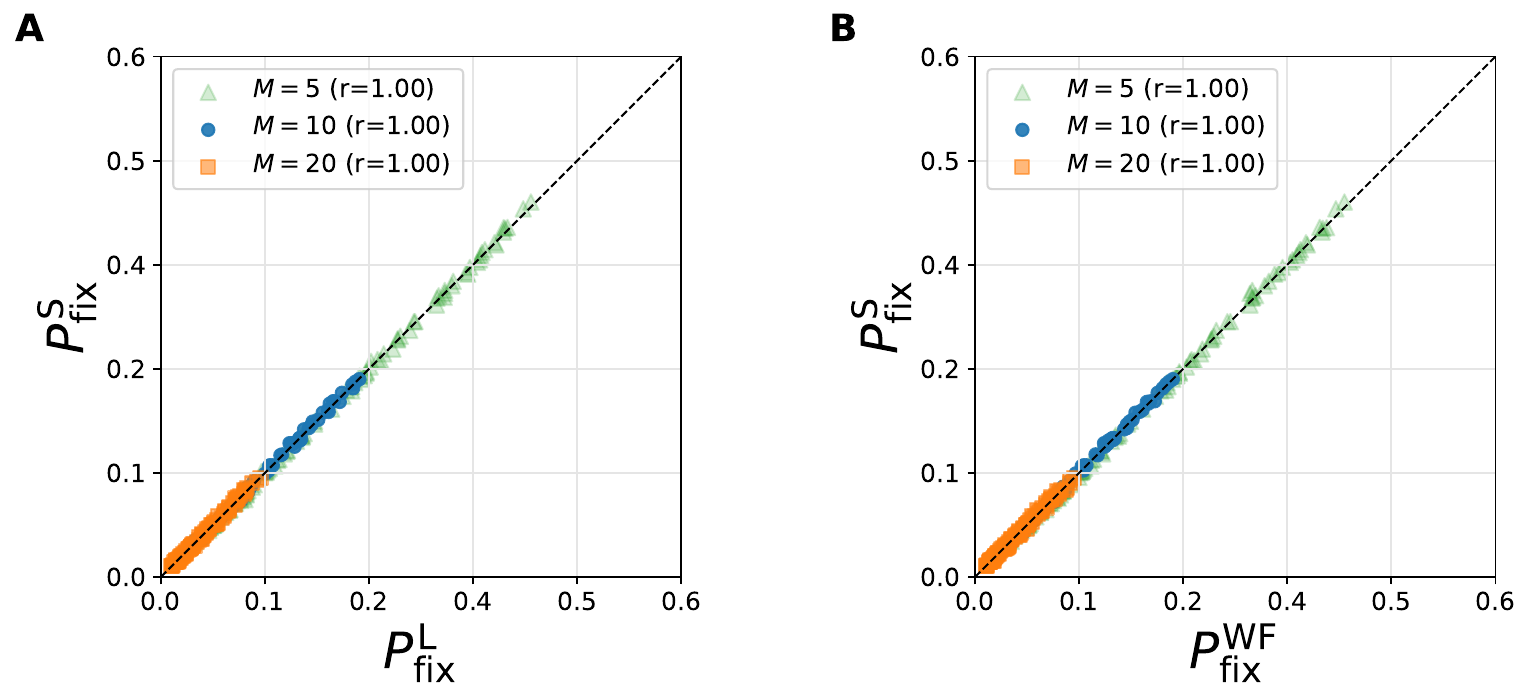}
\caption{ \textbf{Comparison of the series expansion and numerical fixation probabilities in populations with multiple co-evolving alleles.}
A: Selection coefficients were sampled from a uniform distribution in the $[0.00,0.02)$ range, with $20$ independent realizations for $M=5$, $10$ independent realizations for $M=10$,
and $5$ independent realizations for $M=20$ (the population size was set to $N=100$ in all cases). For each realization of the selection coefficients, $M$ Langevin fixation probabilities were computed
using $2 \times 10^4$ independent runs; in each run, Eq.~\eqref{Lng:eq:1} was solved numerically using the Euler–Maruyama method.
The simulations used the time step $dt = 1/20N$, where $N=100$ is the population size. The allele was considered extinct or fixed once its frequency reached $1/4N$ or $1 - 1/4N$, respectively.
The initial allele frequencies are given by $x^i_0 = 1/M,~\forall i$. Shown is a scatter plot of all fixation probabilities computed by series expansion, $P_\fix^{\mathrm{S}}$ (Eq.~\eqref{eq:main-expansion} with $k_\text{max} = 36$), vs.
Langevin fixation probabilities, $P_\fix^{\mathrm{L}}$. The dashed black line is $y=x$; linear correlation coefficients $r$ are computed separately for each value of $M$.
B: Selection coefficients were sampled as in (A). For each realization of the selection coefficients, $M$ Wright-Fisher fixation probabilities were computed
using $2 \times 10^4$ independent runs, each run resulting in the fixation of one of the $M$ alleles.
The initial allele frequencies are given by $x^i_0 = 1/M,~\forall i$. Shown is a scatter plot of all fixation probabilities computed by series expansion, $P_\fix^{\mathrm{S}}$ (Eq.~\eqref{eq:main-expansion} with $k_\text{max} = 36$), vs.
Wright-Fisher fixation probabilities, $P_\fix^{\mathrm{WF}}$. The dashed black line is $y=x$; linear correlation coefficients $r$ are computed separately for each value of $M$.
}
\label{fig:LWFmultcomparo}
\end{figure}

\begin{figure}[!htbp]
\centering
\includegraphics[width=1.05\linewidth]{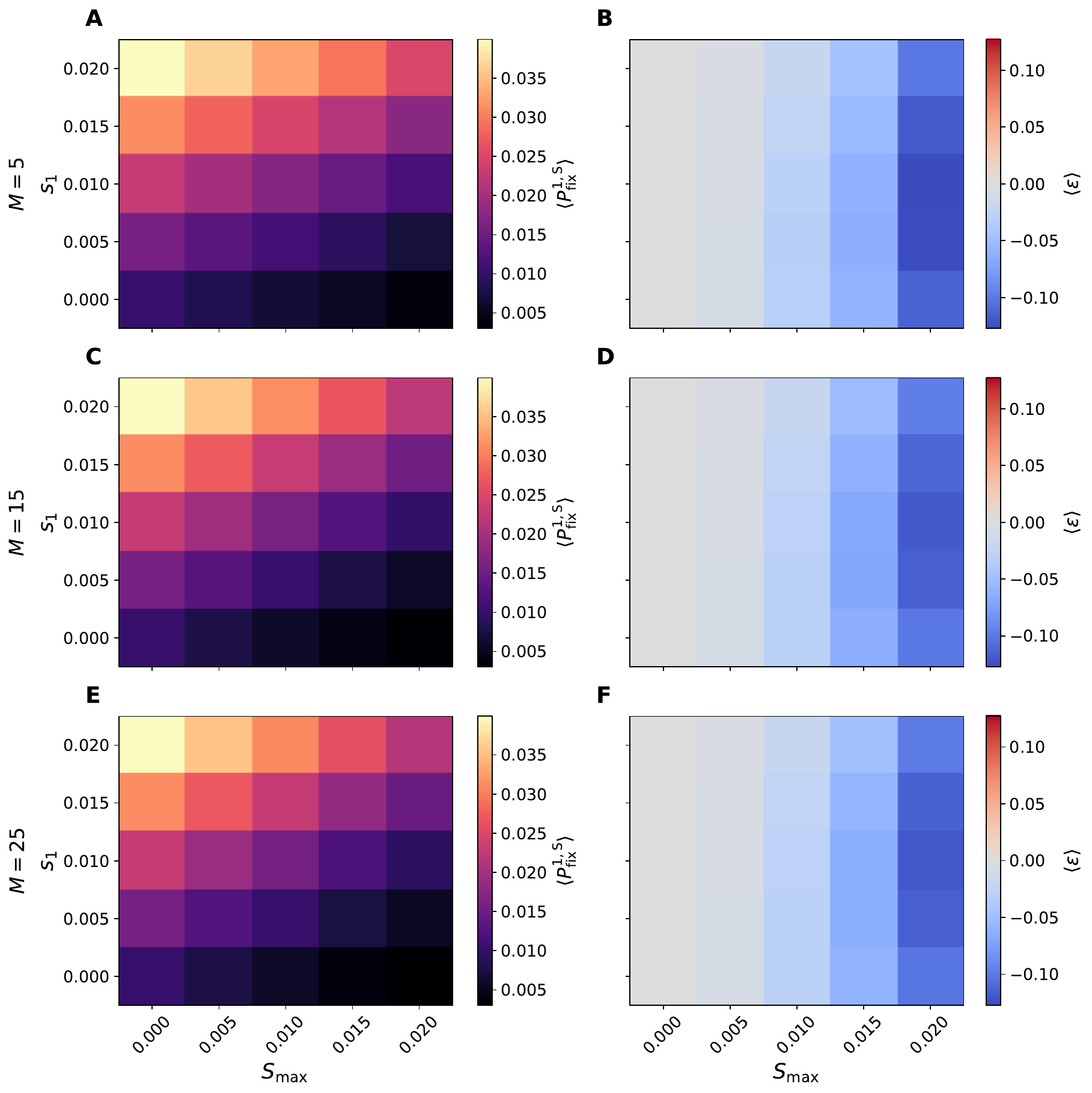}
\caption{ \textbf{Fixation probabilities of the target mutation in the presence of interfering beneficial mutations.}
Fixation probabilities of the target mutation with the selection coefficient $s_1$, $P_\fix^{1,\mathrm{S}}$, were computed using Eq.~\eqref{eq:main-expansion} with $k_\text{max} = 36$, for $5$ systems with $M$ alleles and
$s_1 = \{0.000, 0.005, 0.010, 0.015, 0.020\}$.
The initial allele frequencies were set to $x^1_0 = 1/N$, $x^i_0 = (1 - 1/N)/(M -1),~i = 2, \dots, M$; the population size is $N=100$.
For each value of $s_1$, selection coefficients $s_2, \dots, s_{M-1}$ were uniformly sampled in the $[0.00, S_{\max})$ range, with $s_M = 0.0$
as required by our formalism. For each combination of $s_1$ and $S_{\max}$, the $P_\fix^{1,\mathrm{S}}$ values in panels (A), (C), (E) were averaged over $100$ independently
sampled sets of selection coefficients $s_2, \dots, s_{M-1}$. Panels (A), (C), (E) show these average fixation probabilities, $\langle P_\fix^{1,\mathrm{S}} \rangle$, for $M = 5, 15, 25$, respectively.
Panels (B), (D), (F) show relative errors between the exact fixation probabilities and the mean-field Kimura approximation, Eq.~\eqref{Eq:Kimura:eff}:
$\varepsilon = (P_\fix^{1,\mathrm{S}} - P_\fix^{1,\mathrm{K}})/P_\fix^{1,\mathrm{S}}$, where $P_\fix^{1,\mathrm{K}}$ denotes Kimura's fixation probability.
For each combination of $s_1$ and $S_{\max}$, the relative errors were averaged over $100$ independent realizations of selection coefficients used in panels (A), (C), (E).
}
\label{fig:Tinter2}
\end{figure}

\begin{figure}[!htbp]
\centering
\includegraphics[width=1.05\linewidth]{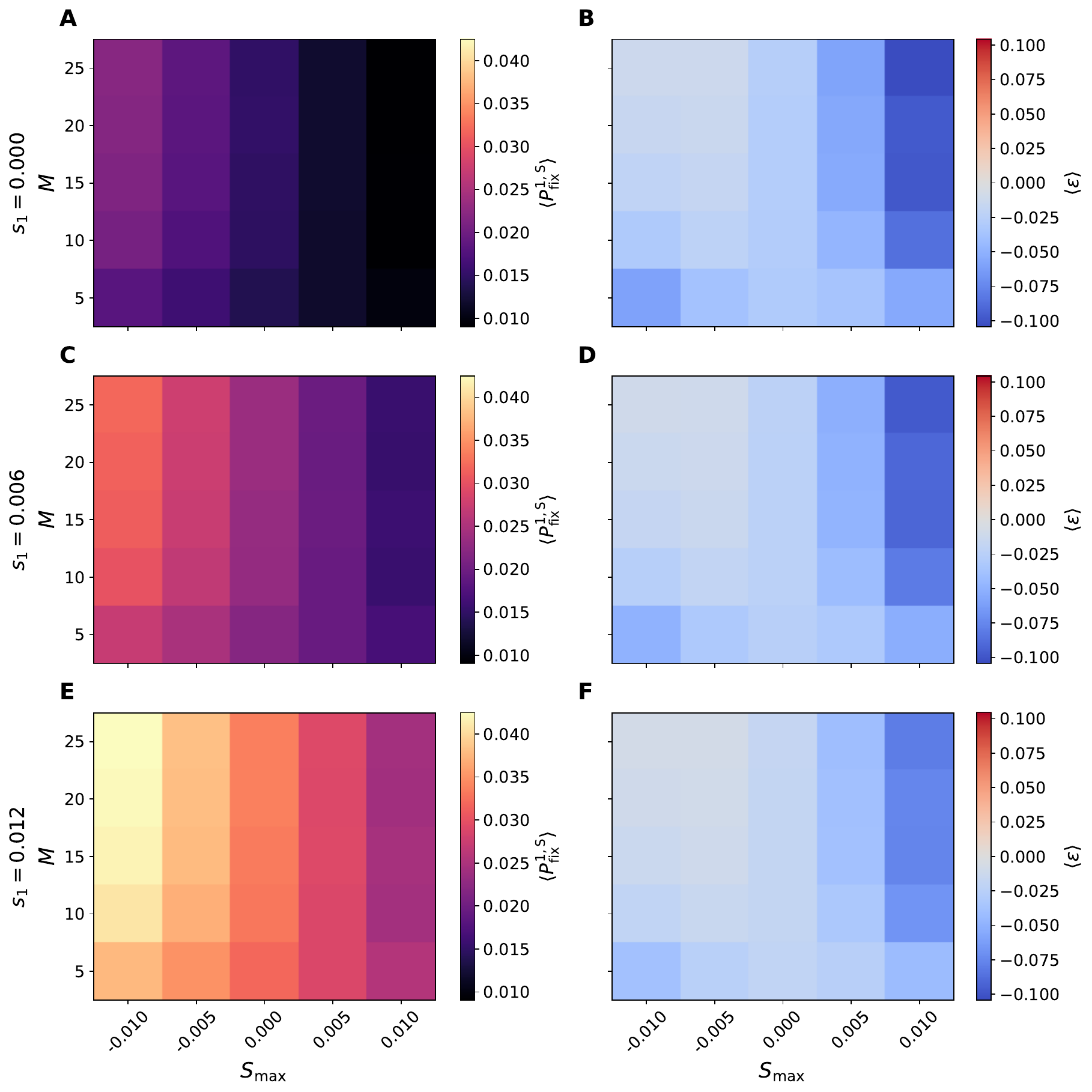}
\caption{ \textbf{Fixation probabilities of the target mutation in the presence of interfering deleterious mutations.}
Fixation probabilities of the target mutation with the selection coefficient $s_1$, $P_\fix^{1,\mathrm{S}}$, were computed using Eq.~\eqref{eq:main-expansion} with $k_\text{max} = 36$, for $5$ systems with
$M = \{5,10,15,20,25\}$.
The initial allele frequencies were set to $x^1_0 = 1/N$, $x^i_0 = (1 - 1/N)/(M -1),~i = 2, \dots, M$; the population size is $N=100$.
For each value of $M$, selection coefficients $s_2, \dots, s_{M-1}$ were uniformly sampled in the $[-0.01, S_{\max})$ range, with $s_M = 0.0$
as required by our formalism. For each combination of $M$ and $S_{\max}$, the $P_\fix^{1,\mathrm{S}}$ values in panels (A), (C), (E) were averaged over $100$ independently
sampled sets of selection coefficients $s_2, \dots, s_{M-1}$. Panels (A), (C), (E) show these average fixation probabilities, $\langle P_\fix^{1,\mathrm{S}} \rangle$, for $s_1 = 0.000, 0.006, 0.012$, respectively.
Panels (B), (D), (F) show relative errors between the exact fixation probabilities and the mean-field Kimura approximation, Eq.~\eqref{Eq:Kimura:eff}:
$\varepsilon = (P_\fix^{1,\mathrm{S}} - P_\fix^{1,\mathrm{K}})/P_\fix^{1,\mathrm{S}}$, where $P_\fix^{1,\mathrm{K}}$ denotes Kimura's fixation probability.
For each combination of $M$ and $S_{\max}$, the relative errors were averaged over $100$ independent realizations of selection coefficients used in panels (A), (C), (E).
}
\label{fig:Tinter3}
\end{figure}

\begin{figure}[!htbp]
\centering
\includegraphics[width=1.05\linewidth]{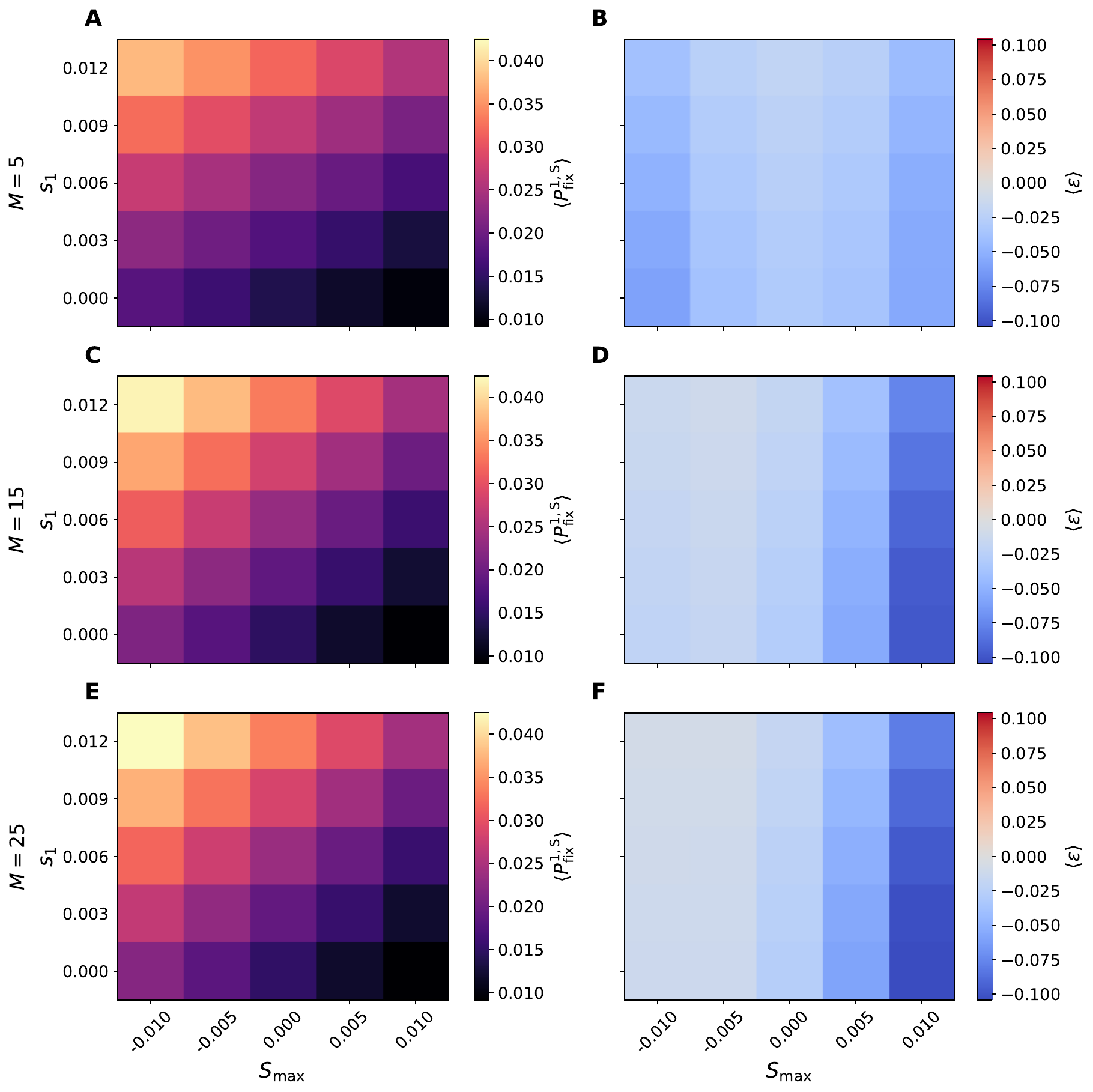}
\caption{ \textbf{Fixation probabilities of the target mutation in the presence of interfering deleterious mutations.}
Fixation probabilities of the target mutation with the selection coefficient $s_1$, $P_\fix^{1,\mathrm{S}}$, were computed using Eq.~\eqref{eq:main-expansion} with $k_\text{max} = 36$, for $5$ systems with $M$ alleles and
$s_1 = \{0.000, 0.003, 0.006, 0.009, 0.012\}$.
The initial allele frequencies were set to $x^1_0 = 1/N$, $x^i_0 = (1 - 1/N)/(M -1),~i = 2, \dots, M$; the population size is $N=100$.
For each value of $s_1$, selection coefficients $s_2, \dots, s_{M-1}$ were uniformly sampled in the $[-0.01, S_{\max})$ range, with $s_M = 0.0$
as required by our formalism. For each combination of $s_1$ and $S_{\max}$, the $P_\fix^{1,\mathrm{S}}$ values in panels (A), (C), (E) were averaged over $100$ independently
sampled sets of selection coefficients $s_2, \dots, s_{M-1}$. Panels (A), (C), (E) show these average fixation probabilities, $\langle P_\fix^{1,\mathrm{S}} \rangle$, for $M = 5, 15, 25$, respectively.
Panels (B), (D), (F) show relative errors between the exact fixation probabilities and the mean-field Kimura approximation, Eq.~\eqref{Eq:Kimura:eff}:
$\varepsilon = (P_\fix^{1,\mathrm{S}} - P_\fix^{1,\mathrm{K}})/P_\fix^{1,\mathrm{S}}$, where $P_\fix^{1,\mathrm{K}}$ denotes Kimura's fixation probability.
For each combination of $s_1$ and $S_{\max}$, the relative errors were averaged over $100$ independent realizations of selection coefficients used in panels (A), (C), (E).
}
\label{fig:Tinter4}
\end{figure}

\begin{figure}[!htbp]
\centering
\includegraphics[width=1.05\linewidth]{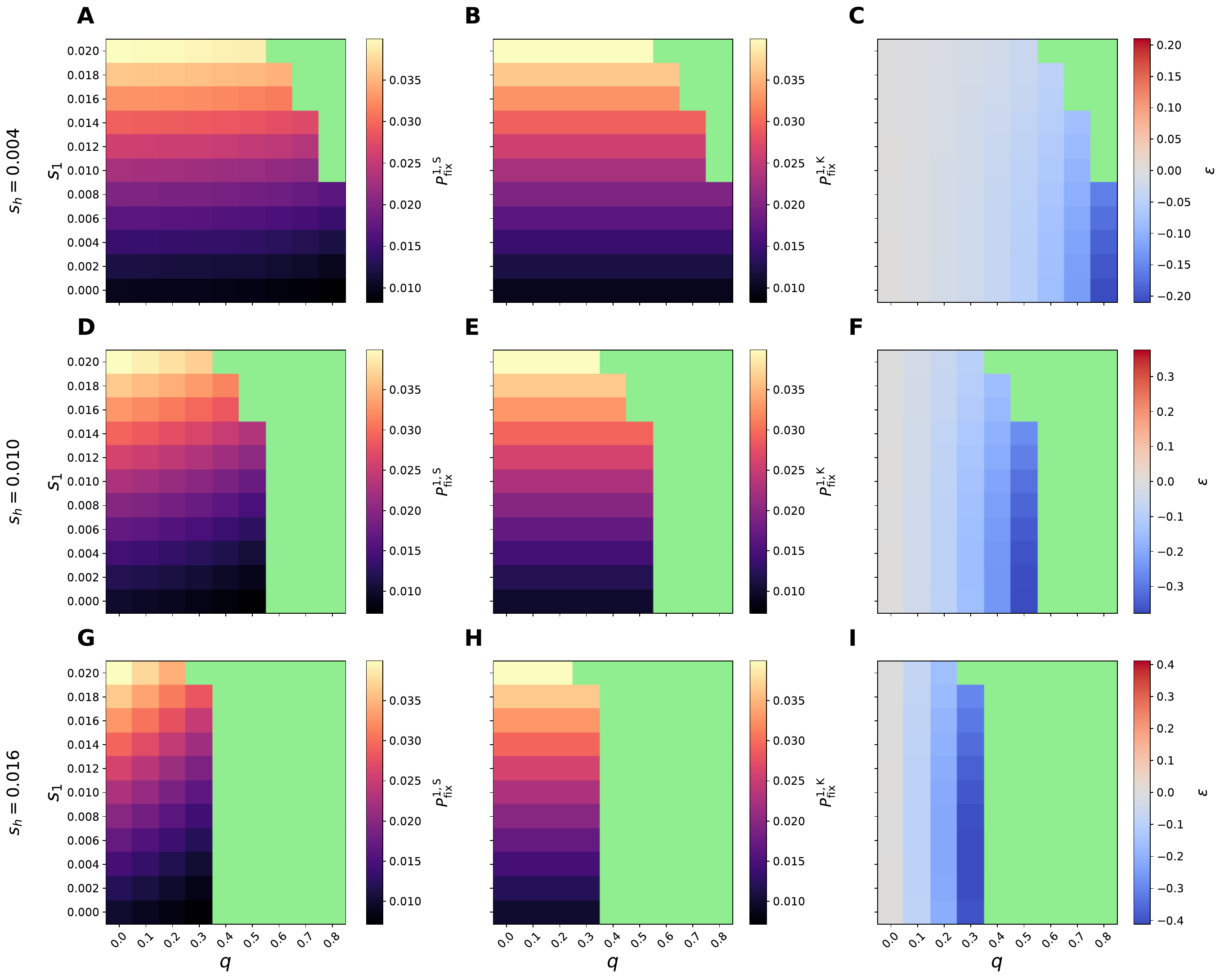}
\caption{ \textbf{Fixation probabilities of a target mutation in the presence of interfering beneficial and deleterious mutations with zero mean.}
Fixation probabilities $P_\fix^{1,\mathrm{S}}$ of the target mutation with the selection coefficient $s_1$ were computed using Eq.~\eqref{eq:main-expansion} with $k_\text{max} = 36$.
Kimura fixation probabilities $P_\fix^{1,\mathrm{K}}$ were computed using Eq.~\eqref{Eq:Kimura:eff} with $\bar{s}_\text{bkg} = 0$.
The initial allele frequencies were set to $(x^1_0, x^2_0, x^3_0) = (1/N, (1 - 1/N)q, (1 - 1/N)(1 - q))$.
The corresponding selection coefficients are $(s_1 + {q s_h}/{(1 - q)}, {s_h}/{(1 - q)}, 0)$; the population size is $N=100$.
Panels (A),(D),(G) show heatmaps of $P_\fix^{1,\mathrm{S}}$ as a function of $q$ and $s_1$, for $s_h = (0.004, 0.010, 0.016)$, respectively.
Panels (B),(E),(H) show the corresponding heatmaps of $P_\fix^{1,\mathrm{K}}$.
Panels (C),(F),(I) show the relative errors between series expansion and Kimura fixation probabilities,
$\varepsilon = (P_\fix^{1,\mathrm{S}} - P_\fix^{1,\mathrm{K}})/P_\fix^{1,\mathrm{S}}$, for each value of $q$ and $s_1$.
Green areas indicate parameter combinations where at least one of the selection coefficients is greater than $0.025$ in magnitude,
preventing reliable convergence of the series expansion formula.
}
\label{fig:qscan2}
\end{figure}

\end{document}